\documentclass[fleqn,usenatbib]{mnras}

\usepackage{newtxtext,newtxmath}
\usepackage[T1]{fontenc}

\DeclareRobustCommand{\VAN}[3]{#2}
\let\VANthebibliography\thebibliography
\def\thebibliography{\DeclareRobustCommand{\VAN}[3]{##3}\VANthebibliography}

\usepackage{graphicx}	% Including figure files
\usepackage{amsmath}	% Advanced maths commands
\usepackage{amssymb}	% Extra maths symbols
\usepackage{booktabs}
\usepackage{lipsum}
\newcommand{\cc}{cm$^{-3}$}
\newcommand{\Msun}{M$_{\odot}$}

\newcommand{\Isolated}{\texttt{Isolated}}
\newcommand{\FaceOn}{\texttt{FaceOn}}
\newcommand{\EdgeOn}{\texttt{EdgeOn}}
\newcommand{\NoCR}{\texttt{NoCR}}
\newcommand{\ADV}{\texttt{ADV}}
\newcommand{\DIF}{\texttt{DIF}}

\newcommand{\ryChange}[1]{{\color{black}{#1}\color{black}}}

\title[Impact of Cosmic Rays on Ram Pressure Stripping]{Stress-Testing Cosmic Ray Physics: The Impact of Cosmic Rays on the Surviving Disk of Ram Pressure Stripped Galaxies}

\author[Farber, Ruszkowski, Tonnesen \& Holguin]{
Ryan J. Farber,$^{1}$\thanks{E-mail: rjfarber@umich.edu}
Mateusz Ruszkowski,$^{1}$
Stephanie Tonnesen,$^{2}$
Francisco Holguin,$^{1}$
\\
$^{1}$Department of Astronomy, University of Michigan, 1085 S. University Ave., Ann Arbor, MI 48109, USA\\
$^{2}$Center for Computational Astrophysics, Flatiron Institute, 162 5th Avenue, New York, NY 10010, USA;
}

\date{Accepted XXX. Received YYY; in original form ZZZ}

\pubyear{2021}

\begin{document}
\label{firstpage}
\pagerange{\pageref{firstpage}--\pageref{lastpage}}
\maketitle

\begin{abstract}
Cluster spiral galaxies suffer catastrophic losses of the cool, neutral gas component of their interstellar medium due to ram pressure stripping, contributing to the observed quenching of star formation in the disk compared to galaxies in lower density environments. However, the short term effects of ram pressure on the star formation rate and AGN activity of galaxies undergoing stripping remain unclear. Numerical studies have recently demonstrated cosmic rays can dramatically influence galaxy evolution for isolated galaxies, yet their influence on ram pressure stripping remains poorly constrained. We perform the first cosmic-ray magneto-hydrodynamic simulations of an $L_{*}$ galaxy undergoing ram pressure stripping, including radiative cooling, self-gravity of the gas, star formation, and stellar feedback. We find the microscopic transport of cosmic rays plays a key role in modulating the star formation enhancement experienced by spirals at the outskirts of clusters compared to isolated spirals. Moreover, we find that galaxies undergoing ram pressure stripping exhibit enhanced gas accretion onto their centers, which may explain the prevalence of AGN in these objects. In agreement with observations, we find cosmic rays significantly boost the global radio emission of cluster spirals. Although the gas removal rate is relatively insensitive to cosmic ray physics, we find that cosmic rays significantly modify the phase distribution of the remaining gas disk. These results suggest observations of galaxies undergoing ram pressure stripping may place novel constraints on cosmic-ray calorimetry and transport.
\end{abstract}


%%%%%%%%%%%%%%%%% BODY OF PAPER %%%%%%%%%%%%%%%%%%

\section{Introduction}
Galaxies exhibit strikingly different properties depending upon the local density of their environment \citep{dressler1980galaxy}. Spirals inhabiting high density environments such as clusters (``cluster spirals'') tend to be redder, more anemic in neutral hydrogen gas, have lower star formation rates, and stronger magnetic fields than their more isolated counterparts in the field \citep{hubble1931velocity,butcher1978b,boselli2006environmental}. 

Multiband observations of relatively isolated star-forming spirals orbiting in cluster environments ubiquitously detect copious amounts of multiphase gas pointing away from cluster center in `tail' structures extending up to 100 kpc (e.g., in X-rays, \citealt{sun200670,sun2009spectacular}; in H$\alpha$, \citealt{zhang2013narrow}; in 21-cm emission \citealt{oosterloo2005large}; in CO, \citealt{jachym2017molecular}; \citealt{moretti2018gasp}; and see the GASP survey for more examples \citealt{poggianti2017gaspSurvey}). 

The relative motion between the intracluster medium (ICM) and the galaxy's interstellar medium (ISM) can lead to gas removal from the disk if the ram pressure (RP), $P_{\rm ram} = \rho v^2$, exceeds the gravitational restoring force per unit area, $P_{\rm grav}$ \citep{gunn1972infall,roediger2009ram}. This ram pressure `stripping' (RPS) should lead to eventual quenching of star formation with most of the gas expected to be stripped on the first infall \citep{jaffe2015budhies}.

However, observations of cluster spirals undergoing RPS can detect moderate \textit{enhancements} of star formation \citep{vulcani2018enhanced}. This star formation enhancement may be related to observations of high efficiency neutral to molecular gas conversion in RPS galaxies \citep[][]{moretti2020gasp,moretti2020high}, suggesting compression-induced star formation. Moreover, recent observations by the GASP survey suggest the AGN fraction is strongly enhanced relative to the field \citep{poggianti2017agn,radovich2019gasp}.

% ST:
Some hydrodynamical simulations have studied the effect of RP on the surviving gas disk.  \citet{schulz2001multi} coined the term ``disk annealing", or the compression of the inner surviving disk via angular momentum transport. \citet{tonnesen2009gas} found that at low RP strengths, more gas was compressed than removed by the ICM, and \citet{Tonnesen2019} argued that early compression could have a long-lasting impact on the amount of gas that is stripped.

%ST: 
Additional simulations have focused on the impact of RP on the star formation rate (SFR) of galaxies, although no consensus has been reached. While many simulations predict that in some cases RP can cause a modest increase in the disk SFR \citep[][]{kronberger2008influence,steinhauser2016simulations,ruggiero2017fate,lee2020dual}, some find that the global SFR is suppressed \citep[][]{tonnesen2012star,roediger2014star,lee2020dual}, and others find that in a few cases the SF can be boosted by a factor of several \citep[][]{bekki2013galactic}. These simulations span a range of galaxy masses, RP strengths, and numerical implementations, and therefore cannot yet be combined into a coherent picture. 

%SKT: 
A growing body of literature has \ryChange{explored} the impact of magnetic fields on ram pressure stripped galaxies \ryChange{using numerical simulations}. While magnetohydrodynamic (MHD) simulations have shown that draping of a magnetized ICM \citep{lyutikov2006magnetic,ruszkowski2007impact,ruszkowski2008cosmic,dursi2008draping,pfrommer2010detecting,Ruszkowski2014} helps to reproduce the smooth (as opposed to clumpy) morphology of RPS tails in agreement with recent observations \citep[][]{muller2021highly}, they find little impact on the gas stripping rate compared to purely hydrodynamic simulations \citep{Ruszkowski2014}. However, \citet{Ruszkowski2014} find an increase in magnetic pressure along the disk due to draping. Similarly, simulations with disk magnetic fields find little impact on the stripping rate as long as the gas surface density is not impacted \citep{Tonnesen2014,ramos2018mhd}. Interestingly, \citet{ramos2018mhd} find magnetized, flared disks act to deflect ICM material towards the galactic center region.    

In contrast, the effect of cosmic rays (CR) on RPS remains largely unexamined. \citet{bustard2020cosmic} performed simulations including CR, radiative cooling, and the derived star formation history of the Large Magellanic Cloud. They showed cosmic-ray driven galactic winds in combination with RPS can contribute to the Magellanic Stream. To our knowledge, CR have never been studied for more massive L$_{*}$ galaxies in a cluster environment. Yet CR have shown to play a crucial role in the evolution of isolated galaxies \citep[][see \citealt{zweibel2017basis} for a review]{ensslin2007cosmic,everett2008milky,uhlig2012galactic,booth2013simulations,salem2014cosmic,simpson2016role,girichidis2016launching,girichidis2018cooler,pfrommer2017simulating,pfrommer2017gamma,Wiener2017,Wiener2019,ruszkowski2017global,Farber2018,butsky2018role,heintz2018parker,holguin2019role,Chan2019,ji2020properties,hopkins2020but,hopkins2021testing,semenov2021cosmic}.

Including CR is essential to understanding the physics underlying observations of RPS galaxies. The nonthermal pressure of CR tends to produce disks of larger scale-height, which should be more easily stripped. Likewise, CR driven galactic winds tend to be cooler and higher density than thermal outflows, suggesting more efficient removal of the neutral medium from galaxies. On the other hand, previous work has shown CR suppress the star formation rate in galaxies, as their nonthermal pressure counteracts contraction of gas to star-forming densities. Cosmic ray models with consequently reduced stellar feedback may exhibit weaker outflows, diminishing the amount of gas that is stripped.

Beyond the dynamics of RPS, CR are fundamental towards understanding radio continuum observations of RPS galaxies, the measurements of which indicate a global radio excess compared to the far-infrared (FIR) to radio correlation (FRC) \citep{dickey19841,beck1988,yun2001,paladino2006}. Since enhanced star formation would boost both the radio and the FIR \citep[][]{lacki2010}, previous work suggested magnetic compression by the ICM wind on the leading edge of the galaxy could explain the boosted radio emission \citep{scodeggio1993}.

However, \citet{murphy2009} utilized \textit{Spitzer} FIR and VLA radio continuum imaging to show a \textit{paucity} of radio emission on the leading edge of the galaxy's orbital motion, precisely where one would expect magnetic compression to dominate. Nevertheless, they observe global radio enhancement. The local radio deficits with global radio enhancement was confirmed by \citet{vollmer2009,vollmer2010,vollmer2013} in Virgo cluster galaxies with multiwaveband measurements. They propose that both the local deficits and global enhancement of radio emission can be explained by variations in the cosmic-ray electron number density.\footnote{The radio emission is expected to be produced via synchrotron emission as the CR electrons gyrate along magnetic field lines.} Although the magnetic field is compressed on the leading edge, cosmic-ray electrons may be easily stripped and their consequently low density suppresses the radio emission. Meanwhile, shocks driven into the ISM by the interaction with the ICM can re-accelerate CR, boosting the global radio emission. However, \citet{pfrommer2010detecting} suggest magnetic draping can explain the deficits in radio emission.

In this work, we determine the impact of CR and their transport, on properties of galaxies undergoing RPS. The outline of this paper is as follows: In \S \ref{sec:Methods} we describe the initial conditions, boundary conditions, galaxy model, and numerical methods utilized to perform this work including magnetohydrodynamics (MHD), CR, radiative cooling, self-gravity of the gas, star formation, and stellar feedback. In \S \ref{sec:Results} we present and discuss our results. In \S \ref{subsec:Morphology} we consider the morphological evolution of the simulations we performed, finding CR crucially modify the outflows in isolated galaxies, yet do not evidently modify the morphology when galaxies undergo RPS. Thus, the stripping rates analyzed in \S \ref{subsec:Stripping} do not show much difference with or without CR. However, in \S \ref{subsec:PhaseSpace} we find cosmic rays protect low-temperature gas from being stripped, possibly helping to explain observations of molecular gas in RPS tails. Cosmic rays dramatically influence star formation, as we show in \S \ref{subsec:StarFormation}. Intriguingly, we find in \S \ref{subsec:AccretionGalacticCenter} that CR modify the accretion rate onto the galactic center with important implications for observations of AGN in RPS galaxies. The transport of CR fundamentally impacts the radio emission, which we discuss in \S \ref{subsec:Radio}. We indicate limitations of our study and directions for future work in \S \ref{subsec:Caveats}. Finally, we conclude in \S \ref{sec:Conclusions}.

% ===============================================
% ===============================================
% ===============================================
% ===============================================
% ===============================================

\section{Methods}
\label{sec:Methods}
\subsection{Numerical Techniques} \label{subsec:NumericalTechniques}
We performed our simulations using the adaptive mesh refinement MHD code FLASH 4.2.2 \citep{Fryxell2000,Dubey2008}, with the directionally unsplit staggered mesh (USM) solver \citep{Lee2009,Lee2013}. The USM is a finite volume, high order Godunov scheme that utilizes constrained transport to satisfy the solenoidal constraint of Maxwell's equations to machine precision.  

Additionally, we include self-gravity, radiative cooling, star formation and feedback as source and sink terms in the MHD equations. We include the passive advection of a concentration variable $C$, used to mark the initial disk gas\footnote{We set C = 1 in the disk and zero elsewhere.}. We further extend the MHD equations to include CR as a second ultrarelativistic fluid (\citealt{yang2012fermi}; \citealt{yang2013fermi}; \citealt{yang2017spatially}; \citealt{ruszkowski2017global}; \citealt{holguin2019role}); see \citet{Farber2018} for the system of equations we solve.

% ===================== *** SELF GRAVITY *** ==========================
To include self-gravity of the baryons (gas and stellar population particles\footnote{We utilize static potentials to include the gravitational influence of pre-existing stars; stellar population particles form during the simulation runtime.}) we solve the Poisson equation using the Huang \& Greengard multigrid solver in FLASH \citep{huang1999fast,ricker2008direct}. The multigrid solver implemented in FLASH extends \citet{huang1999fast} for compatibility with FLASH's numerical structure, namely, finite volume discretization of the fluid equations with shared data on an oct-tree mesh, enabling efficient parallelization. That is, the multigrid method utilizes a direct solver for individual mesh blocks; see \citet{ricker2008direct} for further details.

% ===================== *** Cooling/Heating *** ==========================
We utilized the hybrid scheme for radiative cooling and heating of \citet{gnedin2011environmental}. The implementation automatically switches between an explicit and implicit solver depending on the timestep constraint, enabling efficient and accurate treatment of radiative cooling and heating. For the rates, we interpolate to the nearest temperature and density using a table generated with Cloudy \citep{ferland1998cloudy} assuming a constant solar metallicity and a constant metagalactic UV background; see, e.g., \citet{semenov2021cosmic} for further details.

% ===================== *** SPEED LIMITER *** ==========================
To accelerate the computations we impose a minimum timestep dt$_{\rm min} = 10^{4}$ yr. We do so while maintaining numerical stability by limiting the bulk and generalized sound speeds to

\begin{equation}
    v_{\rm max} \leq C_{\rm cfl} \frac{\Delta x}{\Delta t_{\rm min}}
\end{equation}
\noindent
where $C_{\rm cfl} = 0.2$ is the Courant-Friedrichs-Lewy number and $\Delta x$ is the width of a cell.

We impose the speed limit via dissipation of thermal and CR energy such that the generalized sound speed obeys the speed limit $c_{s} = \sqrt{(\gamma_{\rm th} p_{\rm th} + \gamma_{\rm cr} p_{\rm cr}) / \rho}$ and similarly for the bulk speed.\footnote{\ryChange{Note that ensuring dt$_{\rm min} = 10^{4}$ yr is equivalent to a ceiling of $\sim$10$^4$\,km\,s$^{-1}$ at our best resolution. Since the maximum inflow speed is about ten times smaller, we don't expect our simulations to be impacted by the bulk speed limiter.}} Rather than limiting the Alfven speed directly, we utilize the hybrid Riemann solver implemented in FLASH, as modified to utilize the HLLC Riemann solver in smooth regions and the LLF Riemann solver in shock-detected regions for increased numerical stability (Dongwook Lee, priv. comm., 2020).

% ===================== *** RESOLUTION/REFINEMENT *** ==========================
We perform our simulations in a box of dimensions (-32 kpc, 32 kpc)$^{3}$ with the galaxy centered at the origin with the spin axis pointing in the $z$-direction. We uniformly resolve |z| < 4 kpc with 7 levels of refinement such that our galactic disk achieves a resolution of 127 pc and resolution degrades away from the galactic disk to a base grid with 4 levels of refinement and a physical resolution of $\sim$1 kpc. 

Note that our box is relatively small to minimize the resolution elements covering |z| < 4 kpc; we adopt such a step-wise refinement pattern to avoid ``corners'' of different resolution elements, which we found produced spuriously reflected waves and generated unphysical structures in the magnetic field.

% ===================== *** BOUNDARY CONDITIONS *** ==========================
We employ diode boundary conditions (modified to prevent inflow while including self-gravity, see Appendix \ref{sec:AppTrueDiodeBCs}) on all box faces. For the \FaceOn\ and \EdgeOn\ runs, after waiting 40 Myr to allow the initial conditions to relax, we inject a wind through the -z and -y boundaries, respectively, following the parameters of \citet{Ruszkowski2014}. That is, we model an ICM wind with a density of n$_{\rm wind} = 5 \times 10^{-4}$\,\cc, temperature T$_{\rm wind} = 7 \times 10^7$\,K, magnetic field strength of 2\,$\mu$G perpendicular to the wind, zero along the direction of the wind, and a maximum wind speed of 1300 km/s. We use an accelerating profile of the wind speed to model the orbital motion of the galaxy falling towards the center of the cluster \citep{Tonnesen2019}, following the model of \citet{Ruszkowski2014}, namely $v_{w}(t) = f(t) v_{\rm max,w}$ where

\begin{equation}
    f_{\rm in}(t) = 1 - 
    \begin{cases}
      1 - 1.5 x^2 + 0.75x^3 & \text{if} x \leq 1 \\
      0.25(2 - x)^3 & \text{if} 1 < x < 2\\
      0 & \text{if} x \geq 2
    \end{cases}
\end{equation}
with $x \equiv t/\Delta t$ and $\Delta t \approx 59$\,Myr. That is, the wind reaches maximum velocity at 158\,Myr (since we delay onset of the wind 40\,Myr) and is constant thereafter. See Table \ref{tab:NumericalParameters} for a list of the parameters employed in this study.
\begin{table}
\begin{center}
\caption{\label{tab:NumericalParameters}Numerical Parameters.}
\begin{tabular}{ c c }
\hline
 Name & Value \\ 
 \hline

$dt_{min}$           & $10^{4}$ yr                          \\
$C_{\rm cfl}$        & $0.2$                             \\

$\rho_{*, \rm min}$  & 1 \cc                             \\
$E_{\rm SN}$         & 10$^{51}$ erg                       \\
$f_{\rm th}$         & 0.717                             \\
$f_{\rm cr}$         & 0.1                               \\

$\kappa_{\parallel}$ & 3 $\cdot 10^{28}$ cm$^2$ s$^{-1}$ \\
$\kappa_{\perp}$     & 0                                 \\

$M_{*,\rm disk}$     & 10$^{11}$ \Msun                   \\
$M_{*,\rm bulge}$    & 10$^{10}$ \Msun                   \\
$M_{\rm gas,disk}$   & 10$^{11}$ \Msun                   \\ 
$M_{\rm DM,core}$    & 1.1 $\cdot$ 10$^{11}$ \Msun       \\  
$r_{\rm 0,*,disk}$   & 4 kpc                             \\
$r_{\rm 0,gas,disk}$ & 7 kpc                             \\
$r_{\rm 0,*,bulge}$  & 0.4 kpc                           \\
$z_{\rm 0,*,disk}$   & 0.25 kpc                          \\
$z_{\rm 0,gas,disk}$ & 0.4 kpc                           \\

R$_{\rm disk}$       & 26 kpc                            \\
B$_{\rm 0,x}$        & 1 $\mu$G                          \\
a$_{\rm 0}$          & 10$^3$                            \\

$\rho_{\rm 0,CGM}$   & 9.2 $\times 10^{-5}$ \cc          \\
T$_{\rm 0,CGM}$      & 4.15 $\times 10^{6}$ K            \\

n$_{\rm wind}$    & 5 $\times 10^{-4}$ \cc            \\
T$_{\rm wind}$       & 7 $\times 10^{7}$ K               \\
v$_{\rm max,wind}$   & 1300 km/s                         \\
B$_{\perp,wind,1}$   & 2 $\times 10^{-6}$G               \\
B$_{\perp,wind,2}$   & 2 $\times 10^{-6}$G               \\
B$_{\parallel,wind}$ & 0                                 \\

L$_{\rm box}$        & 64.8 kpc                          \\
dx$_{\rm min}$       & 127 pc                            \\
dx$_{\rm max}$       & 1 kpc                             \\
|z|$_{\rm dx,min}$   & 4 kpc                             \\

\hline
\end{tabular}
\end{center}
\end{table}

% ===================== *** STAR FORMATION *** ==========================
\subsubsection{Star Formation and Feedback \label{subsubsec:StarFormAndFeedback}}
We employ the star formation and stellar feedback prescriptions used in the ART \citep{kravtsov1999phd,kravtsov2002constrained,rudd2008effects,gnedin2011environmental} simulations of galaxy evolution \citep{semenov2016nonuniversal,semenov2017physical,semenov2018galaxies,semenov2021cosmic}. That is, we permit gas to form stars when the gas density exceeds a critical value $n_{*, \rm min}$ = 1\,\cc\ with the star formation rate density $\rho_{*}$ parameterized to match the Kennicutt-Schmidt relation \citep{schmidt1959rate,kennicutt1998global}:

\begin{equation}
    \dot{\rho_{*}} = \epsilon_{\rm ff} \frac{\rho}{t_{\rm ff}}
\end{equation}
where $\epsilon_{\rm ff}$ is the star formation efficiency per free-fall time $t_{\rm ff}$. We set $\epsilon_{\rm ff}$ = 0.01 in agreement with observationally inferred low local star formation and long galactic depletion times of star-forming gas (see, \citealt{krumholz2007slow,leroy2017cloud,semenov2018galaxies}).

When the gas density of a cell exceeds the minimum density for star formation we create a stellar population particle with the mass proportional to $N$, where $N$ is the number of occurrence, drawn from a Poisson distribution, characterized by the expected value $\lambda = \rho_{*} dV dt / m_*,min$. However, we limit the value of $N$ to not exceed $m_{\rm gas}/m_{\rm *,min}$, where $m_{\rm *,min} = 10^4$\,\Msun\ is the minimum stellar population mass we enforce to avoid creating a computationally intractably large number of particles. We also require the total stellar population particle mass not to exceed $2/3$ $m_{\rm gas}$ to avoid consuming all the gas in the cell. Upon creation of the stellar population particle, we remove its mass from the gas in the cell it inhabits.

% ===================== *** FEEDBACK *** ==========================
For 40\,Myr immediately succeeding the creation of a stellar population particle we apply feedback from massive stars, modeling proto-stellar jets, massive stellar winds and radiation pressure, preceding type II supernovae. We sample our stellar population particles with a Chabrier IMF to determine the contribution from massive stars $\gtrsim$\,8\,\Msun. For each massive star we inject 0.1 $\times 10^{51}$\,erg as cosmic-ray energy, as well as thermal energy and momentum according to the subgrid model of \citet{martizzi2015supernova}. 

The \citet{martizzi2015supernova} momentum feedback subgrid model takes into account local conditions and our resolution to inject the appropriate amount of momentum produced during the (unresolved) Sedov-Taylor phase (see also \citealt{walch2015energy,kim2015momentum,iffrig2015mutual}). We boost the injected momentum by a factor of 5 to account for the unresolved clustering of supernovae \citep{gentry2017enhanced,gentry2019momentum}, the influence of CR on the supernova momentum deposition \citep[][]{diesing2018effect} and to account for advection errors \citep{agertz2013toward,semenov2018galaxies}.

% ===================== *** COSMIC RAYS *** ========================
\subsubsection{Cosmic Ray Models \label{sec:CosmicRayModels}}
We bracket the parameter space of CR transport and calorimetry via three cases: (1) No CR (\NoCR), i.e., modeling the case of complete calorimetry. (2) CR that simply advect with the thermal gas (\ADV). This case models unresolved tangled structure of the magnetic field in the galactic disk, preventing CR from escaping into the circumgalactic medium (CGM) or effectively a strong suppression of cosmic-ray diffusion and cooling near star formation sites \citep{semenov2021cosmic}. (3) Cosmic rays anisotropically diffuse along magnetic field lines with a diffusion coefficient parallel to the magnetic field of $3 \times 10^{28}$ cm$^2$ s$^{-1}$ and zero perpendicular diffusion (\DIF). We note that these three cases are meant to bracket the possible results of more detailed CR transport modeling, which remains highly uncertain (see the dozens of models of \citealt{hopkins2021testing}).\footnote{That is, \ADV\ can represent the limiting case of very slow CR transport, \NoCR\ models the limit of very fast transport, and the diffusion coefficient adopted for \DIF\ is a moderate value that should be near the peak wind driving efficincy \citep[][]{salem2014cosmic} and is motivated by models of cosmic-ray propagation (see \citealt{grenier2015nine} and references therein).}

\subsection{Galaxy Model \label{GalaModeSec}}
We model a massive spiral galaxy with a flat rotation curve $v_{\rm disp} = 200$\,km/s, composed of a gaseous disk, hot halo, stellar disk, stellar bulge, and dark matter halo initially in hydrostatic equilibrium (see \citealt{tonnesen2009gas}; \citealt{tonnesen2010tail}).

To directly follow \citet{roediger2006ram} we use a Plummer-Kuzmin potential for the stellar disk \citep{miyamoto1975three}, a Hernquist profile for the stellar bulge \citep{hernquist1993n}, and a \citet{burkert1995structure} potential for the dark matter halo \footnote{This potential is consistent with observed rotation curves \citep{trachternach2008dynamical}.} (see Table \ref{tab:NumericalParameters} for the masses and scale lengths of each component).

We employ the same magnetic field configuration initial conditions as model TORL of \citet{Tonnesen2014}. Namely, we initialize the magnetic field to be weak in the galactic central region (where the velocity field changes rapidly), peaking in strength a few kpc from the galactic center, diminishing gradually with increasing galactic radius, and set to zero outside the disk. That is, we employ the following vector potential (with $A_z$ set to a constant outside the disk)

\begin{equation}
    A_x = A_y = 0
\end{equation}

\begin{equation}
    A_z = \sqrt{a_{zf}} e^{-6 R_{\rm cyl}} \frac{ -6 \sin(2.5 R_{\rm cyl}) - 2.5 \cos(2.5 R_{\rm cyl})  }{6^2 + 2.5^2}
\end{equation}

\begin{equation}
    a_{zf} = 1000 (-|z| + 1)^{80}
\end{equation}
Note that the cutoff in magnetic field strength at the disk-edge ensures the magnetic pressure is subdominant to the thermal pressure, reducing disk expansion as well as growth of the magnetic field due to shear between the disk and the CGM. 

The plasma beta, which is the ratio of thermal to magnetic pressures, $\beta \equiv P_{\rm gas} / P_{\rm mag}$ ranges from 100 to a maximum of $\sim$2 in the disk midplane a few kpc from the galactic center region (see Figure 2 of \citealt{Tonnesen2014}). Since the magnetic pressure is subdominant to the thermal pressure the magnetic field does not have a strong effect on the disk. The chosen magnetic field morphology enables easier comparison to our previous work \citep{Tonnesen2014} as it is reproducible, reduce variability due to instabilities, and are divergence-free.

% ============================================================
% ============================================================
% ============================================================
% ============================================================
% ============================================================
% ============================================================

\section{Results \& Discussion}
\label{sec:Results}
We begin the presentation of our results in Section \ref{subsec:Morphology} by graphically illustrating the nine runs we have analyzed and indicating their morphological evolution in Figures \ref{fig:SlicePlotEarly} through \ref{fig:SlicePlotLate}. Then we explore the impact of CR on the stripping rates in Section \ref{subsec:Stripping} and the gas phase distribution in Section \ref{subsec:PhaseSpace}. In Section \ref{subsec:StarFormation} we discuss the impact of CR and their feedback on the star formation rate in galaxies undergoing RPS at cluster outskirts. We consider the impact of CR on the accretion of gas toward the galactic center in \ref{subsec:AccretionGalacticCenter}. See Table \ref{tab:Nomenclature} for our nomenclature.
%Finally, we consider a simple proxy for the radio emission in our models in \ref{subsec:Radio}.
\begin{table}
\caption{\label{tab:Nomenclature}Nomenclature.}
\begin{tabular}{ c c }
\hline
 Name & Description \\ 
 \hline

 $f_{\rm th}$ & thermal fraction of SN energy.           \\
 $f_{\rm cr}$ & cosmic-ray fraction of SN energy.        \\
 dx  & cell width.                                       \\
 MHD & Magnetohydrodynamics.                             \\
 USM & Unsplit Staggered Mesh.                           \\
 RPS & Ram Pressure Stripping.                           \\
 SFR & Star Formation Rate.                              \\
 SNe & Supernovae.                                       \\
 ICM & Intra-cluster Medium.                             \\
 CGM & Circum-galactic Medium.                           \\
 \Isolated & No ICM wind.                                \\  
 \FaceOn & ICM wind parallel to galactic spin axis.      \\
 \EdgeOn & ICM wind perpendicular to galactic spin axis. \\
 \NoCR & No cosmic rays.                                 \\
 \ADV & Cosmic rays only advect.                         \\
 \DIF & Cosmic rays advect and diffuse.                  \\
 
\hline
\end{tabular}
\end{table}

% ============================================================
\subsection{Galactic Morphology}
\label{subsec:Morphology}
\begin{figure*}
  \begin{center}
    \leavevmode
    \includegraphics[width=\textwidth]{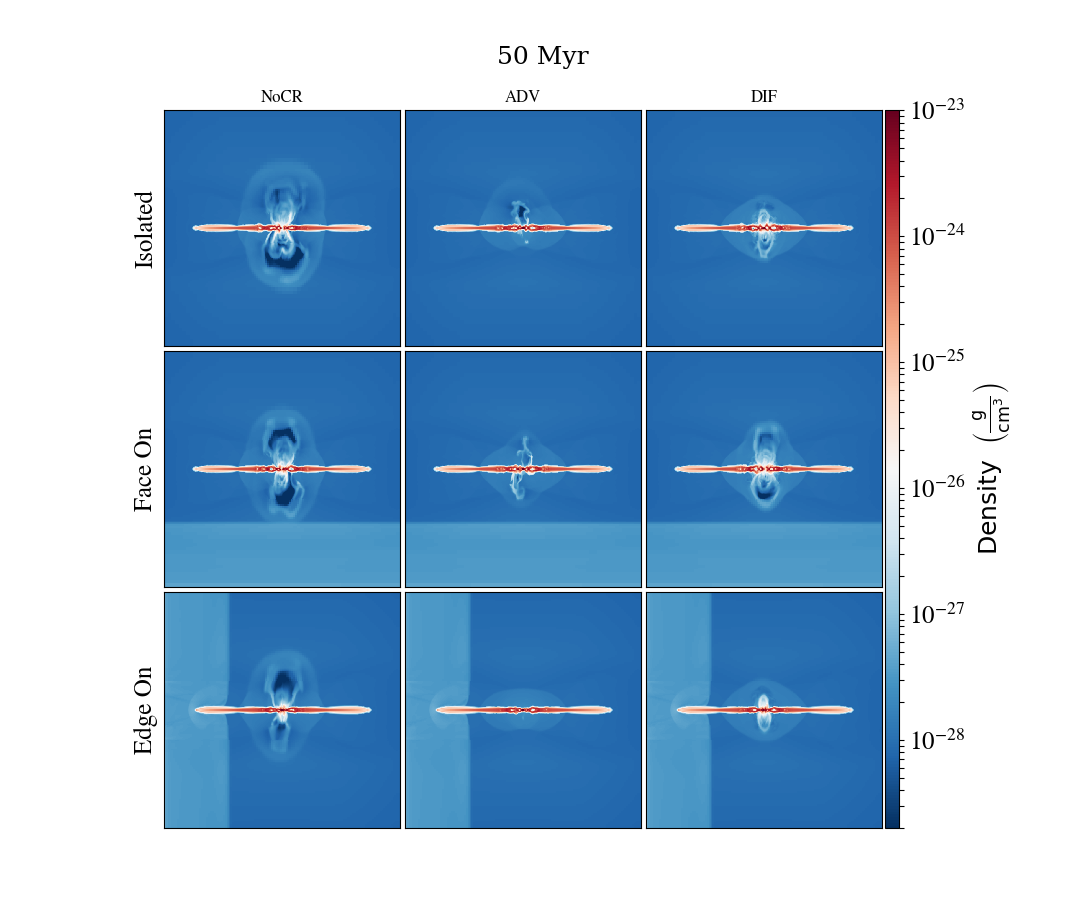}
\caption[]{Slice plots at 50\,Myr, shortly after the galaxies encounter an ICM wind (when appropriate). \NoCR\ (left), \ADV\ (middle), and \DIF\ (right) physics cases are displayed for the \Isolated\ (top), \FaceOn\ (middle), and \EdgeOn\ (bottom) runs. Red colors indicate high density while blue colors indicate low densities.}
\label{fig:SlicePlotEarly}
\end{center}
\end{figure*}

In this section we analyze the morphological evolution of galactic disks. We performed a grid of nine simulations, composed of three CR models and three galaxy models. We performed simulations (a) without CR (\NoCR), (b) where CR purely advect with the gas (\ADV), and (c) including the anisotropic diffusion of CR along magnetic field lines (\DIF\; see Section \ref{sec:CosmicRayModels} for further details). For each CR model we simulated (i) field galaxies not subject to any ICM wind (\Isolated), and galaxies falling into a cluster such that the spin axis points (ii) face-on (\FaceOn) or (iii) orthogonal to (\EdgeOn) the direction of the orbital motion.

In all cases, the gas distribution collapses towards the midplane as a result of radiative cooling removing thermal pressure support. The resulting high densities at the midplane allow stars to form, whose feedback begins to launch a nuclear outflow (see Section \ref{subsubsec:StarFormAndFeedback} for further details on our star formation and feedback methods). After 40 Myr the star formation and feedback cycle is well underway yet still quite similar for all physics cases (see Figure \ref{fig:SFR}). At this time we turn on the ICM wind (for non-\Isolated\ runs). At 50\,Myr (see Figure \ref{fig:SlicePlotEarly}) the ICM wind has just begun to interact with the \FaceOn\ \& \EdgeOn\ runs. At this time, the \NoCR\ runs have the most extended outflow structure while the \ADV\ runs have a relatively weak wind.

\begin{figure*}
  \begin{center}
    \leavevmode
    \includegraphics[width=\textwidth]{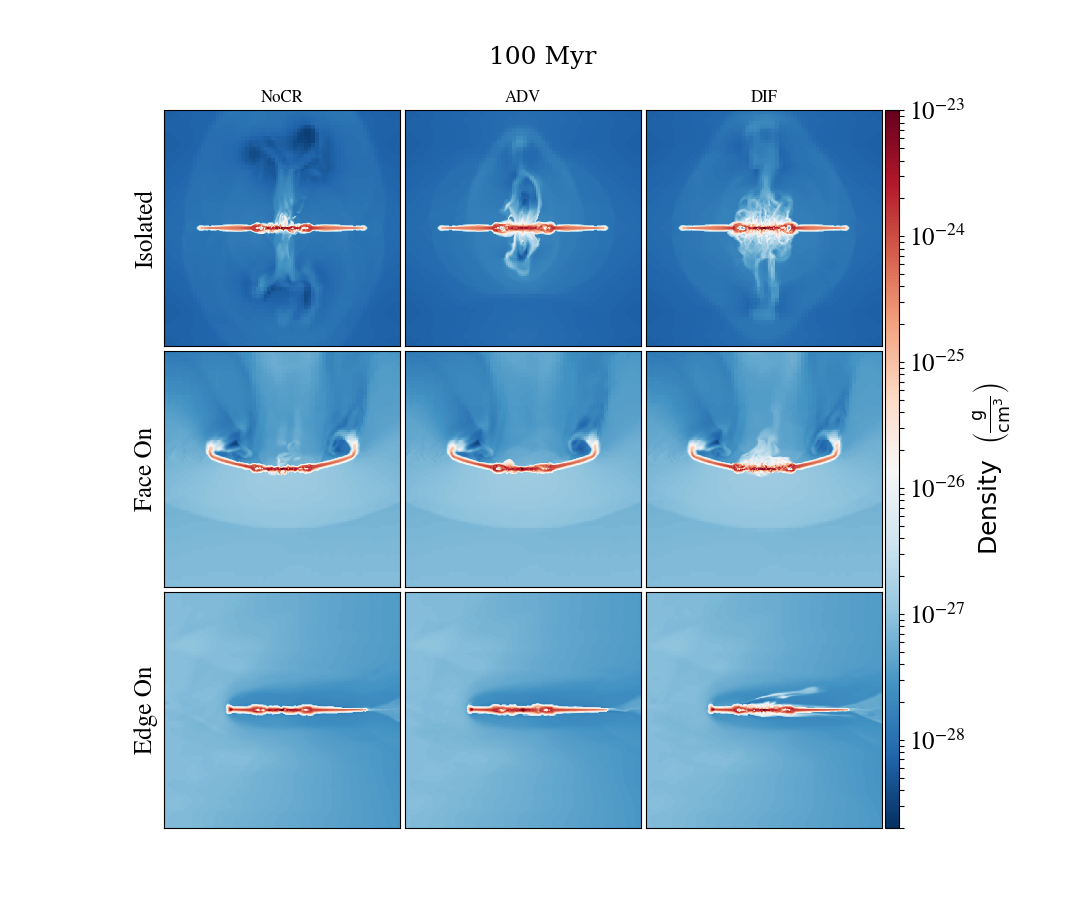}
\caption[]{Same as Figure \ref{fig:SlicePlotEarly} but at 100\,Myr. Galactic winds have developed for the \Isolated\ galaxies while for \FaceOn\ disk gas is bowed backward due to ram pressure. The galactic wind material is absent in the case of \EdgeOn\ as it is readily stripped.}
\label{fig:SlicePlotMiddle}
\end{center}
\end{figure*}

As expected from the classical Gunn \& Gott model for RPS, the ensuing ISM-ICM interaction proceeds from the outermost radii of the disk, where gas is beginning to be stripped by 100\,Myr (see Figure \ref{fig:SlicePlotMiddle}). The \Isolated\ runs have now developed more extended galactic wind structures into the CGM. Note the \DIF\ run has developed a frothy higher density galactic wind structure compared to the relatively low density ``mushroom'' ejecta of the \NoCR\ case, while the weaker galactic wind in the \ADV\ case has not propagated as far into the CGM as \NoCR\ or \DIF. The higher density outflow driven with CR is consistent with previous models of isolated galaxies with cosmic-ray feedback (see \citealt{girichidis2018cooler}). 

Meanwhile, cases with an ICM wind are beginning to diverge from their \Isolated\ galaxy counterparts. Most noticeable is the higher density of the ICM ($\rho \sim 10^{-27}$\,g \cc) compared to the CGM ($200 \rho_{\rm crit} \sim 10^{-28}$\,g \cc). In the \FaceOn\ runs the development of a bowshock $\sim$20\,kpc upstream of the disk is quite evident. Moreover, RPS has begun to affect the disks: the outermost portion of the disks are slightly bowed downstream from the RP of the colliding ICM wind. 

For the \EdgeOn\ runs the disks are slightly pushed downstream at 100\,Myr (see Fig. \ref{fig:SlicePlotMiddle}), with notably absent galactic wind bubbles -- stripping of galactic wind material has been ``caught-in-the-act'' for the \DIF\ case. Otherwise the galaxies in the \EdgeOn\ case do not appear very disturbed.

\begin{figure*}
  \begin{center}
    \leavevmode
    \includegraphics[width=\textwidth]{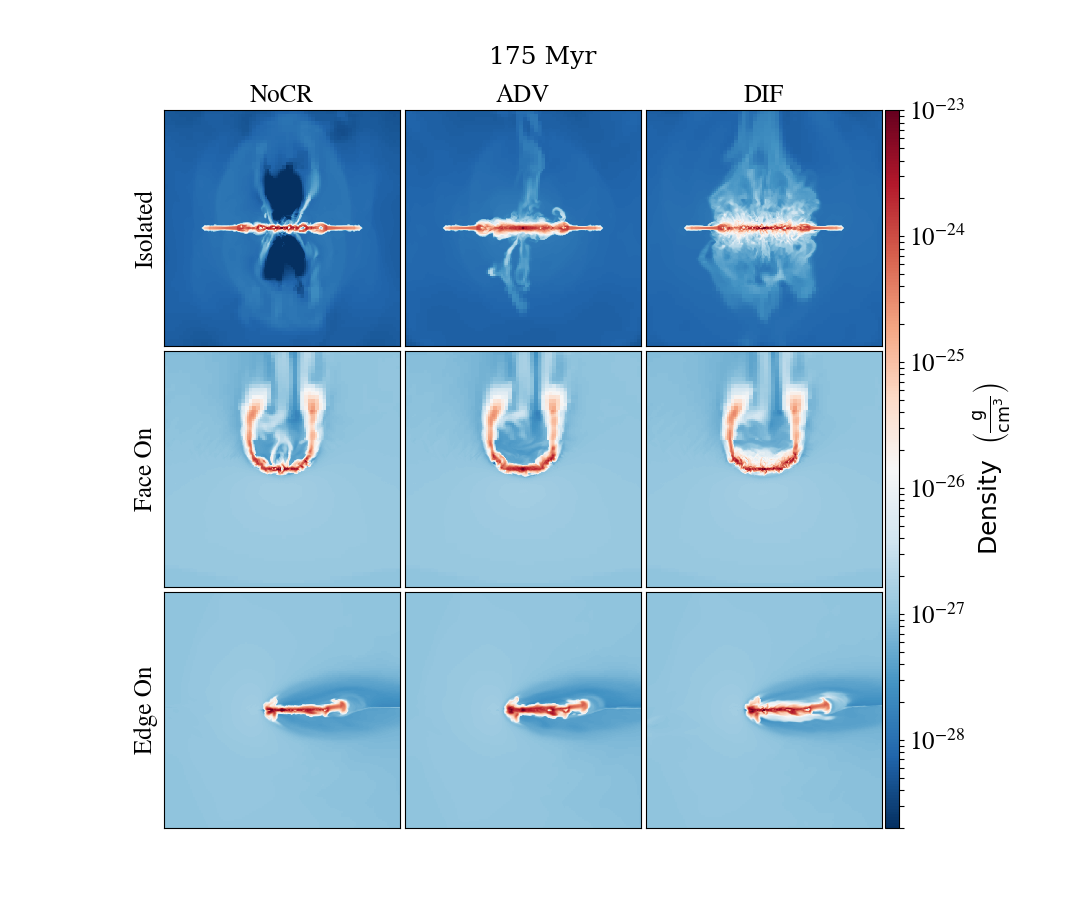}
\caption[]{Same as Figure \ref{fig:SlicePlotEarly} but at 175\,Myr. The \Isolated\ galaxies' winds have continued to expand into the CGM. The \FaceOn\ and \EdgeOn\ galaxies' disks are distorted due to the ram pressure of the ICM wind.}
\label{fig:SlicePlotLate}
\end{center}
\end{figure*}

As the ICM-ISM interaction proceeds, the \FaceOn\ and \EdgeOn\ cases diverge more drastically from the \Isolated\ galaxies. At 175 Myr (see Figure \ref{fig:SlicePlotLate}) the galactic winds of the \Isolated\ runs' have further developed. The \NoCR\ run has evacuated a large scale, biconical, low density ($\rho \sim 10^{-29}$\,g \cc) cavity from its nuclear region into the CGM, in shape reminiscent to the Fermi bubbles \citep[][]{su2010giant}. In contrast, the \DIF\ run has developed a frothy, disk-wide outflow whose relatively high-density ($\rho \sim 10^{-26}$\,g \cc) is in better agreement with observations of the CGM \citep[][]{Werk2013}. Note that stellar feedback similarly occurs to a larger radius in the disk now in the \NoCR\ case as can be noted by the slightly thicker disk compared to the disk outskirts. In the \ADV\ case, the outflow has largely shut off and the gas at large scale heights from the disk is returning in a fountain flow. The main impact of cosmic-ray feedback in this case is the thickening of the gaseous disk due to the nonthermal pressure provided by CR, which efficiently suppresses star formation.

Meanwhile, the pushing of the now maximum RP wind (see Section \ref{subsec:NumericalTechniques} for the acceleration profile) has bowed back the \FaceOn\ disks. Stripping from the ends of the warped disk gas downstream (and out of the computational domain) is evident possibly forming a tail structure (but studying the RPS tail is beyond the scope of this work). The high-density disk gas for the \FaceOn-\ADV\ run is thicker than the other physics cases. Reminiscent of the frothy outflow of the \Isolated\ disk, the \FaceOn-\DIF\ case exhibits a low density skin of gas surrounding the higher density midplane gas as seen in white in our slices. The low density skin possesses a larger scale height downstream, as it is protected by the ``shadow'' of the disk.

While the \FaceOn\ disks are severely warped downstream, the \EdgeOn\ disks appear to exhibit a fairly strong compression on the leading edge, forming a comet-like structure with a distinct high density ``head'' just upstream of (stellar) galactic center and a tail of dense disk gas downstream. Again, the \ADV\ run exhibits a somewhat thicker structure of dense disk gas while the \DIF\ galaxy possess a low density skin, as an outflow (which is rapidly stripped) is driven by diffusing CR.

While we have indicated the most salient differences between our CR physics cases, the most striking observation is how relatively \textit{similar} the morphology is for galaxies subjected to RPS. This is shocking especially after one considers the radical differences of the galactic winds of the \Isolated\ cases. Nevertheless, previous studies have found RPS proceeds at a rate largely unaffected by nonthermal forces (e.g., magnetic fields, see \citealt{Tonnesen2014,Ruszkowski2014}), as suggested by the good agreement between observations and the Gunn \& Gott criterion. In the next section, we investigate the impact of CR on the RPS rate.

% ============================================================
\subsection{Stripping}
\label{subsec:Stripping}
\begin{figure}
  \begin{center}
    \leavevmode
    \includegraphics[width=0.45\textwidth]{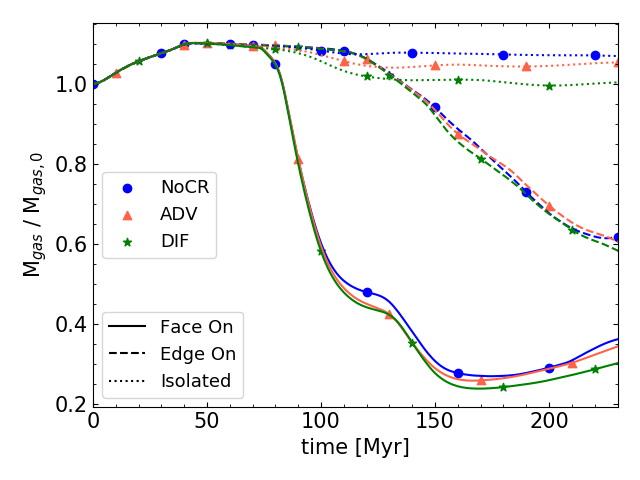}
\caption[]{Evolution of the gas mass in the disk as a function of time. We define disk gas mass to possess a concentration tracer threshold $C > 0.6$ within a cylinder of radius 28.6\,kpc and height $\pm$1\,kpc from the midplane.\footnote{Note the results depend very weakly on the value of $C$ (we considered values ranging from 0.1 - 0.9 to confirm but used 0.6 as in previous work).} Cosmic ray physics cases \NoCR, \ADV, and \DIF\ are indicated as \ryChange{blue circles, red triangles, and green stars}, while the ICM wind types are indicated as solid, dashed, and dotted curves for \FaceOn, \EdgeOn, and \Isolated\ respectively. The stripping rate appears to be fairly insensitive to cosmic ray physics.}
\label{fig:StrippingPlot}
\end{center}
\end{figure}

To determine the impact of CR physics on the RPS rate, we investigate the amount of gas remaining in the disk in our simulations. We define disk gas as possessing a tracer fraction $C > 0.6$ contained within a cylinder of radius 28.6\,kpc and height $\pm$1\,kpc from the disk midplane. 

In Figure \ref{fig:StrippingPlot} we show the mass in the disk as a function of time for our nine simulations. We utilize solid, dashed, and dotted lines to indicate the \FaceOn, \EdgeOn, and \Isolated\ runs while \ryChange{blue circles, red triangles, and green stars} represent \NoCR, \ADV, and \DIF\ physics cases. During the first 40\,Myr the disk mass increases due to radiative cooling induced collapse onto the disk midplane. The disk mass stabilizes until $\sim$75\,Myr when the impact of the ICM wind begins to differentiate the evolution. 

From 75-150\,Myr the \FaceOn\ runs are efficiently stripped of $\sim$80\% of their disk mass. As some stripped material enters the ``shadow'' of the remaining disk and is protected from acceleration to the escape velocity, some material falls back onto the disk from $\sim$175-230\,Myr. In contrast, the \EdgeOn\ runs are more slowly stripped (maintaining $\gtrsim$\,90\% of their initial disk gas mass by 150\,Myr and $\sim$60\% by 230\,Myr) owing to the reduced cross-section for ISM-ICM interaction, in agreement with previous work \citep[][]{roediger2006ram,jachym2009ram}. The \EdgeOn\ runs lack a fallback episode owing to the lack of protective ``shadow.'' 

Meanwhile, the \Isolated\ runs maintain most of their disk mass as expected with the disk gas mass loss due primarily to galactic winds. Interestingly, the \ADV\ run ejects more disk gas mass initially than the \NoCR\ run due to the contrast between the hot, low density cavities driven in the \NoCR\ case compared to the \ADV\ case (see Figure \ref{fig:SlicePlotLate}). Both \ADV\ and \DIF\ runs exhibit periodic fountain flows (see the oscillatory behavior of the dotted lines) or accretion dominating over the outflow, whereas the \NoCR\ disk gas mass is monotonically decreasing.

As anticipated from the largely similar evolution of the \NoCR, \ADV, and \DIF\ cases of the \FaceOn\ and \EdgeOn\ runs (see Figures \ref{fig:SlicePlotEarly}-\ref{fig:SlicePlotLate}), the stripping rates do not differ widely between the physics cases (not more than $\sim$15\%, see Figure \ref{fig:StrippingPlot}). However, knowing that the \Isolated\ runs produce quite distinct galactic wind structures, the fact that CR play a minor role in the removal of gas from RPS galaxies is quite surprising. Nevertheless, we find CR physics does play a role in modifying the temperature-density phase space of cluster spirals, which we discuss next.

\subsection{Gas Phase Distribution in the Disk}
\label{subsec:PhaseSpace}
\begin{figure*}
  \begin{center}
    \leavevmode
    \includegraphics[width=\textwidth]{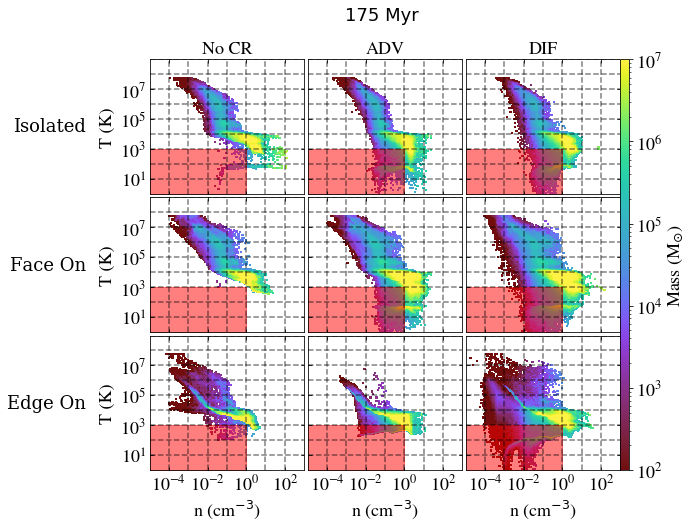}
\caption[]{Gas temperature vs. number density phase plots at 175\,Myr. The left, center, and right columns display \NoCR, \ADV, and \DIF\ physics cases respectively, while the top, middle, and bottom rows show \Isolated, \FaceOn, and \EdgeOn\ ICM wind (or lack thereof) run types. Bright (dark) colors indicate regions with much (little) gas mass. Dashed lines are intended to ease comparison across the various runs. \ryChange{The red highlighted regions of low-temperature and low-density highlight the influence of cosmic rays in allowing such a phase to persist.}}
\label{fig:PhasePlotNT}
\end{center}
\end{figure*}

While the morphology and stripping rates are similar across our three cosmic-ray physics cases (\NoCR, \ADV, and \DIF), examining the galactic disk in detail suggests CR play an interesting role in modifying galactic properties. We start with the phase space distributions of disk\footnote{Note that we select `disk' gas by applying a concentration threshold of $C > 0.6$ and require gas to fit within a cylinder of radius 28.6\,kpc and height $\pm$1\,kpc from the midplane.} gas temperature and density at 175\,Myr in Figure \ref{fig:PhasePlotNT}. The columns from left-right show \NoCR, \ADV, and \DIF\ physics cases respectively, and the rows from top-bottom display \Isolated, \FaceOn, and \EdgeOn\ runs.

In agreement with previous work, the \Isolated\ runs (top row) indicate that including CR permits the presence of low-density, low-temperature gas, as CR provide nonthermal pressure support \citep[][]{ji2020properties,Butsky2020}. Note the low temperature ridge at $T \sim 10^2$\,K \ryChange{is weak in \NoCR\ runs (supported only by magnetic pressure in those cases)}. The low temperature ridge is dimmer (less yellow) for the \DIF\ run related to less CR pressure in the disk supporting that gas. The ridge has a low-mass component extending to lower temperatures in the \DIF\ case related to adiabatic cooling of the galactic wind.

Note that once the wind has impacted the galaxies, we expect that low-density gas will be preferentially removed due to its lower restoring force.  Indeed, in the \NoCR\ run this is clearly seen in both the \FaceOn\ and \EdgeOn\ runs at all temperatures. We also see a small increase in the amount of dense gas, particularly in the \EdgeOn\ run, in agreement with our visual impression of gas compression in Figures 1-3.

However, the gas distribution is somewhat different in the stripped galaxies with CR. While above $\sim$10$^4$\,K, low density gas seems to be removed, at cooler temperatures more low-density gas survives than in the \NoCR\ wind runs \ryChange{(highlighted by the red boxes)}.  We can understand this by the low-temperature ridge in the isolated versus wind runs.  

The low temperature ridge is at slightly lower temperatures when RPS is turned on due to enhanced nonthermal pressures. An elevated magnetic field strength due to compression of the disk is the source of nonthermal pressure in the \NoCR\ case. The low temperature ridge gets brighter when including CR and is brighter for \ADV\ than \DIF\ related to the concentration of CR in the disk. As we will show, CR production is enhanced due to increased SFRs with RP. The low temperature feature extends to lower temperature in the \DIF\ case due to CR outside the disk. Although CR are swept away quickly in the \EdgeOn\ case (as seen in Figure \ref{fig:SlicePlotLate}), they are also constantly replenished by star formation and subsequent feedback. In summary, the inclusion of CRs allows low-temperature, low-density gas to survive in ram pressure disks due to an increase in non-thermal pressure.

The \NoCR\ runs show that RPS efficiently strips low density gas. However, galaxies subjected to RPS (middle and bottom rows) including CR (middle and right columns) contain low density gas that is otherwise stripped in the \NoCR\ runs (left column; bottom left portion of each plot).

Probably the most interesting result is that RPS increases the amount of low-temperature gas and this RPS-induced is effect is stronger when CR are included (see \S \ref{sec:Conclusions}).

% ============================================================
\subsection{Star Formation}
\label{subsec:StarFormation}
\begin{figure}
  \begin{center}
    \leavevmode
    \includegraphics[width=0.49\textwidth]{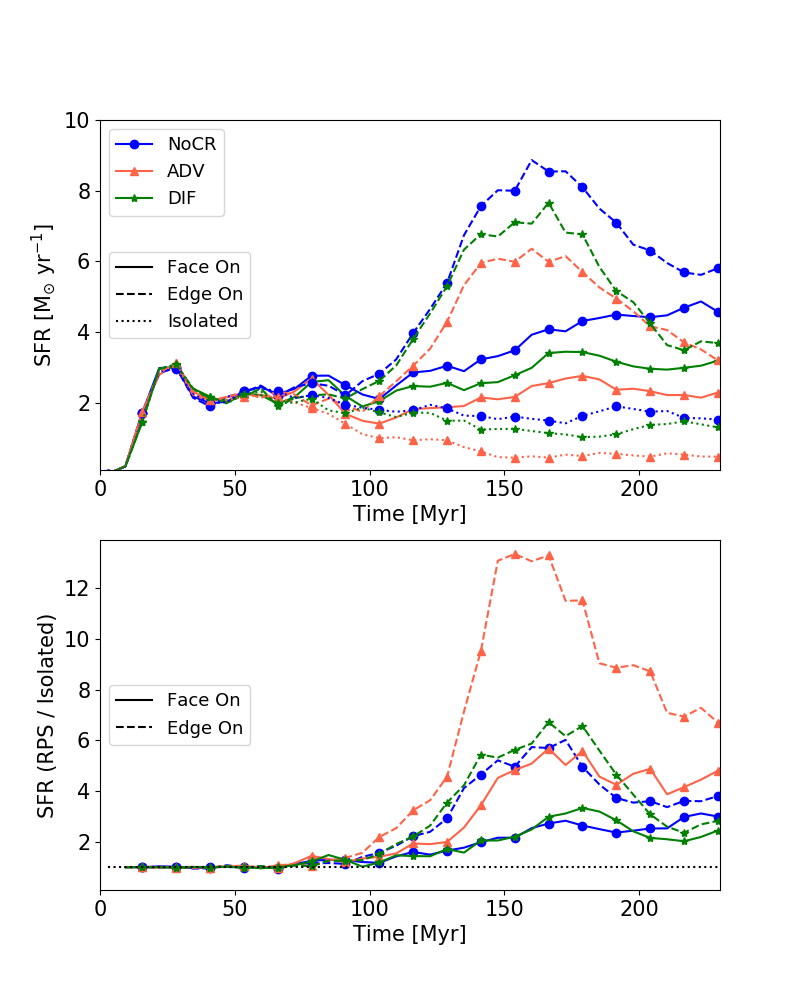}
\caption[]{Time series of SFR (top) and ratio of SFR of the \FaceOn\ (solid) or \EdgeOn\ (dashed) runs to the \Isolated\ (dotted) runs. \NoCR, \ADV, and \DIF\ are marked as \ryChange{blue circles, red triangles, and green stars}, respectively. Clearly the \NoCR\ runs have the highest SFR and \ADV\ receives on average the largest boost in SFR when a galaxy is subjected to an ICM wind.}
\label{fig:SFR}
\end{center}
\end{figure}

Let us begin by examining the physical SFR in the top panel of Figure \ref{fig:SFR}. As previously, we use \ryChange{blue circles, red triangles, and green stars} to respectively indicate \NoCR, \ADV, and \DIF\ cases while solid, dashed, and dotted lines refer to the ICM wind cases of \FaceOn, \EdgeOn, and \Isolated. All physics cases and ICM wind runs have similar SFR up to roughly 100\,Myr. After that point, the ICM-ISM interaction begins to take effect with the SFRs clearly diverging. For all ICM wind runs \NoCR, \DIF, and \ADV\ clearly have the highest to lowest respective SFRs with the \EdgeOn\ runs exhibiting demonstrably higher SFR than \FaceOn\ runs at 175\,Myr \ryChange{(see App. \ref{sec:AppConvergence} for a discussion of convergence)}.

Comparing the SFR of RPS galaxies to their isolated counterparts is more readily achieved looking at ratios in the bottom panel of Figure \ref{fig:SFR}. Again, we observe significant departures in the SFRs of different ICM wind or cosmic-ray physics cases only after 100\,Myr. Subsequently, \FaceOn-\DIF\ and \FaceOn-\NoCR\ show moderate enhancement of SFR over their \Isolated\ counterparts in best agreement with observations ($\sim$ 0.2 dex enhancement of star formation, see \citealt{vulcani2018enhanced}). 

The \FaceOn-\ADV\ run achieves peaks of 4-6 times the value of \Isolated-\ADV\ (which is largely quiescent due to the buildup of cosmic-ray energy). The \NoCR-\EdgeOn\ run similarly initially attains a boost 4 times the \Isolated\ run at 175\,Myr. The \EdgeOn-\ADV\ run shows the most dramatic boost in SFR, exceeding an order of magnitude at $\sim$175\,Myr. Since a fluid dominated by CR has an adiabatic index of 4/3 whereas a thermal plasma has an adiabatic index of 5/3, the CR fluid is more compressible. Thus, when the ICM wind impacts the disk in the \ADV\ run which is dominated by cosmic-pressure, there is a larger increase in the density than the \DIF\ or \NoCR\ cases, explaining the boost in SFR.

Clearly the \EdgeOn-\ADV\ run is inconsistent with the much more modest boosts in star formation found by \citet{vulcani2018enhanced}; in fact, all the \EdgeOn\ runs appear to be inconsistent with observations. \citet{roediger2006ram} found inclination makes a minor difference to RPS until $\gtrsim$ 60$^o$, suggesting most spiral galaxies falling into a cluster can be modeled as \FaceOn. Although both \FaceOn-\DIF\ and \FaceOn-\NoCR\ are consistent with observed SFR enhancements, the degeneracy may be broken when we consider the evolution of mass at the galactic center below.

% ============================================================
\subsection{Feeding the AGN: Accretion onto the Galactic Center}
\label{subsec:AccretionGalacticCenter}
\begin{figure}
  \begin{center}
    \leavevmode
    \includegraphics[width=0.49\textwidth]{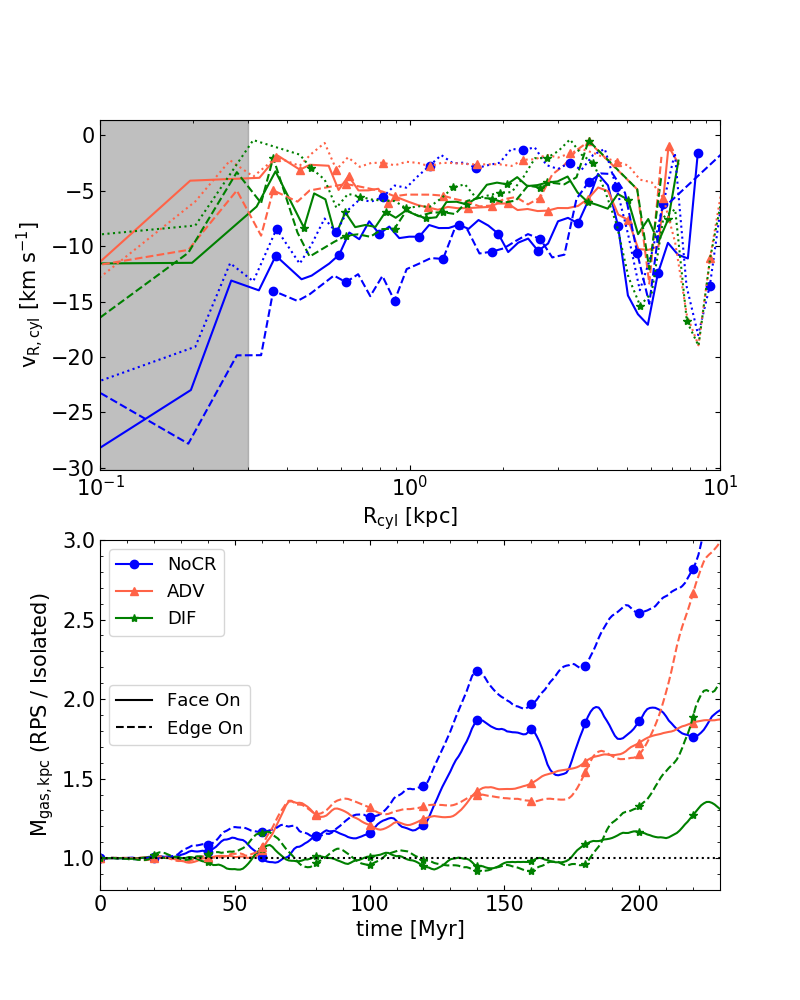}
\caption[]{Top: Time-averaged negative (gas flowing towards the galactic center) cylindrical radial velocity as a function of cylindrical radius. The greyed out region samples less than 10 cells linearly so we caution against over-interpretation. Bottom: galactic center mass ratios of ram pressure stripping runs to \Isolated\ counterparts, in this case defined as a sphere of radius 1\,kpc centered on the galactic center. \NoCR, \ADV, and \DIF\ physics cases are shown as \ryChange{blue circles, red triangles, and green stars} respectively. \FaceOn\ and \EdgeOn\ ratios are indicated by solid and dashed markers respectively.}
\label{fig:Accretion}
\end{center}
\end{figure}

While the SFR plots appear degenerate between the \DIF\ and \NoCR\ models, the galactic accretion rate (Figure \ref{fig:Accretion}) appears to tell a different story. The top panel indicates the accretion rate of material flowing in the disk plane towards the galactic center. That is, we select disk gas (tracer concentration $C > 0.6$ within a cylinder of  radius 28.6\,kpc and height $\pm$1\,kpc from the midplane) and take a one-dimensional profile of the radially inward flowing velocity. We time-average the radial velocity at each radius from 100 - 230\, Myr\footnote{The SFR begins to divrge between the different runs at 100\,Myr. We confirmed our results averaging 50 - 230\,Myr. }. We indicate the \NoCR, \ADV, and \DIF\ models as \ryChange{blue circles, red triangles, and green stars}, with the \FaceOn, \EdgeOn, and \Isolated\ galaxies marked as solid, dashed, and dotted. In general, we see that RPS galaxies have enhanced accretion rates, in qualitative agreement with observations finding RPS galaxies have much higher AGN fractions than galaxies in the field \citep[][]{poggianti2017agn,radovich2019gasp}. Moreover, the runs with CR appear to suppress galactic accretion, as the \NoCR\ run clearly has the strongest accretion rate independent of inclination. Together with the SFR results, this suggests CR must be more weakly coupled to the disk gas. That is, either CR are transported faster or they rapidly experience strong Coulomb and hadronic losses. \ryChange{The accretion rate for the \Isolated\ runs appear to be consistent with recent cosmological models that also find cosmic rays suppress accretion \citep[][]{trapp2021gas}.}

For a clearer picture of the accretion onto the galactic center, we show in the bottom panel of Figure \ref{fig:Accretion} the time evolution of the gas\footnote{\ryChange{Note that the mass converted into stars is relatively modest and does not impact our conclusions.}} mass within a sphere of radius 1\,kpc emanating from the galactic center as a ratio of RPS runs to their \Isolated\ counterparts. \ryChange{For example, we indicate with a blue solid curve the central 1\,kpc gas mass for the ratio of \FaceOn-\NoCR\ to \Isolated-\NoCR.} Until about 100\,Myr, the galactic central mass is largely indistinguishable between ICM wind types and CR physics cases ($\sim$20\% difference). Subsequently, the ISM-ICM interaction drives accretion towards the galactic central region. \ryChange{The nonthermal pressure of CR suppress accretion and as a result, the central gas mass is nearly double for \FaceOn-\NoCR\ compared to \FaceOn-\DIF.} In general, the \NoCR\ runs exhibit the largest central masses followed by the \ADV\ and then \DIF\ runs.

At the end of our simulations, the \DIF\ run has the least mass accreted in the galactic center. This may suggest CR suffer strong $\sim$calorimetric losses, or that the diffusion coefficient must be much faster to reduce the coupling to disk gas, if strong accretion is required to match observations of high AGN fraction in RPS galaxies. However, we note that although weaker, the \DIF\ RPS runs nevertheless have an elevated accreted mass in the galactic center compared to \Isolated\ \ryChange{(that is, the ratio is greater than one)}; it would be interesting if future work can tease out the required accretion to trigger AGN.

% I think this is more appropriate to the results/discussion section 
Note that even in the \Isolated\ runs the galactic center mass increases by a factor of two for cosmic-ray runs and a factor of four for \NoCR\ (not shown here). Our magnetized disk is subject to the magnetorotational instability (MRI; \citealt{balbus1991powerful}) beyond $\sim$4 kpc where the angular velocity begins to diminish radially outward \citep{hawley1999transport}. Since the MRI growth rate is proportional to orbital frequency, growth is strongest for the inner parts of the disk (e.g., $\sim$120\,Myr at 4\,kpc and $\sim$800\,Myr at the disk edge; for more discussion see \citealt{Tonnesen2014}). We also expect gravitational torques (i.e., self-gravity) to transport angular momentum. Thus, we expect some accretion of mass towards the galactic center even for \Isolated. Moreover, since the magnetic field in the disk is strengthened for disks subjected to RPS (see middle panel of Figure \ref{fig:Radio}), the increase in accretion for RPS runs compared to \Isolated\ could be due to more efficient MRI.

The enhanced accretion of \NoCR\ runs over those with CR, particularly over \DIF\ runs, is potentially constraining of CR transport and calorimetry.

% ============================================================

\subsection{Radio Emission}
\label{subsec:Radio}
\begin{figure}
  \begin{center}
    \leavevmode
    \includegraphics[width=0.49\textwidth]{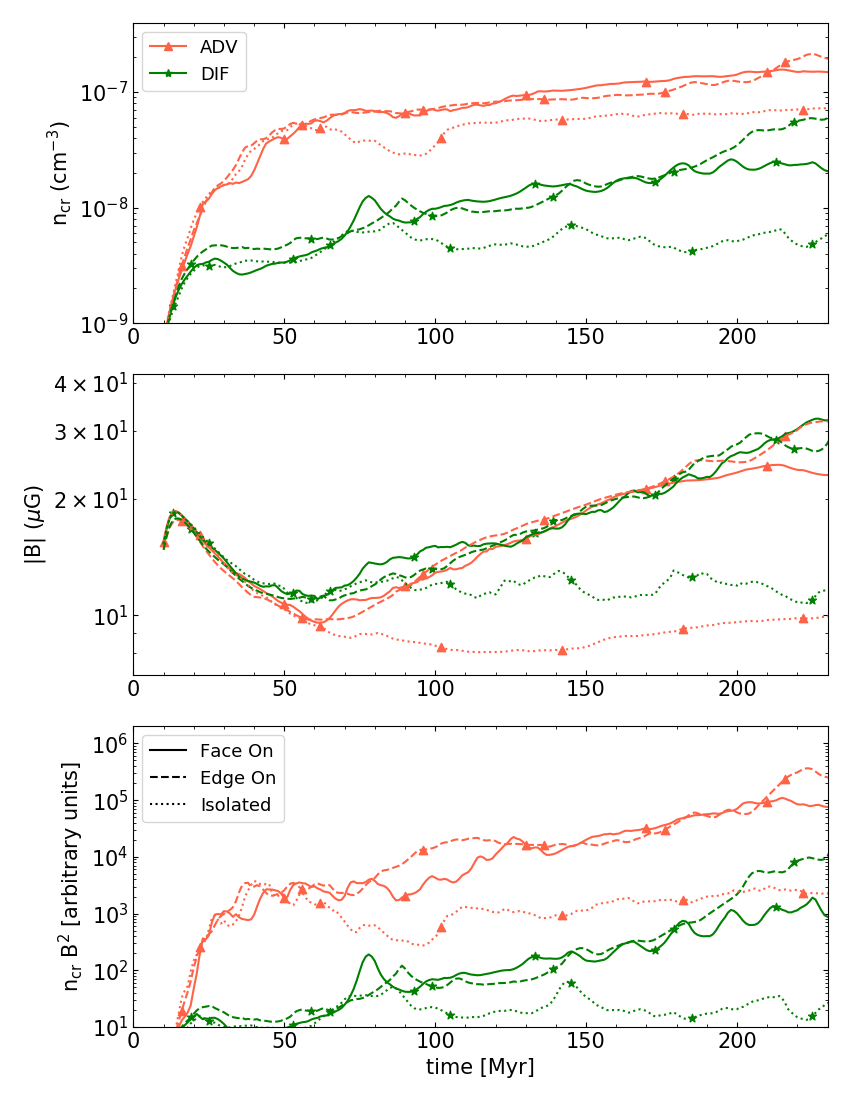}
\caption[]{Time series profiles of cosmic ray number density (top), magnetic field strength (middle), and proxy for radio synchrotron (bottom). \ryChange{red triangles (green stars)} curves indicate \ADV\ (\DIF) physics cases. Solid, dashed, and dotted curves indicate \FaceOn, \EdgeOn, and \Isolated\ runs respectively. RPS both boosts |B| and $n_{\rm cr}$ by a factor of a few, producing a $\sim$2 magnitude boost in expected radio emission.}
\label{fig:Radio}
\end{center}
\end{figure}

RPS galaxies are observed to possess enhanced global radio emission, existing as outliers to the otherwise very tight FRC \citep[][]{murphy2009,vollmer2013}. However, it is unclear if the excess compared to the FRC is due to stronger magnetic fields, elevated CR number densities $n_{\rm cr}$, or both. In Figure \ref{fig:Radio} we determine the relative contribution of these components to a synchrotron radio emission proxy ($n_{\rm cr} B^2$).

In the top panel of Figure \ref{fig:Radio} we see that RPS elevates the $n_{\rm cr}$ by a factor of a few for \ADV\ runs and about one magnitude for \DIF\ runs compared to their \Isolated\ galaxy counterparts, likely related to RPS boosting the SFR (see Figure \ref{fig:SFR}). We highlight that the lower $n_{\rm cr}$ for \DIF\ compared to \ADV\ must be due to the transport of CR out of the galactic disk, as more CR are injected for the \DIF\ run (it has a higher SFR than \ADV). Similarly, the boost in $n_{\rm cr}$ is very similar in the \FaceOn\ and \EdgeOn\ runs even though the boost in SFR is different, again suggesting transport dominates the $n_{\rm cr}$ evolution. Surprisingly, \ADV\ has a smaller boost in $n_{\rm cr}$ despite having a larger boost in SFR than \DIF for runs including RPS. This may suggest fewer CR are stripped in the \DIF\ case. Clearly, exploring the boost in $n_{\rm cr}$ represents an interesting avenue for future research.

Yet the elevated radio emission in RPS galaxies is not only due to the elevated CR number density; in the middle panel of Figure \ref{fig:Radio} we see that the magnetic field strength is similarly boosted by a factor of a few. We also note that the magnetic field strength in the \NoCR runs are boosted by a similar amount. It is somewhat surprising that the magnetic field strength is so similar for all runs with a wind, independent of CR transport or wind orientation. On the one hand, one would expect stronger magnetic field strengths for the \EdgeOn\ runs as they exhibit higher SFRs (see top panel of Figure \ref{fig:SFR}). On the other hand, compression of the disk magnetic field should be more effective for the \FaceOn\ run, given the larger cross-sectional area of the galaxy presented to the ICM wind. We plan on disentangling these effects in future work.

However, since the synchrotron emission depends steeply on the magnetic field strength, the combined effect is a boost of two to three magnitudes, as seen in the bottom panel of Figure \ref{fig:Radio}.

Clearly, CR transport plays a crucial role in driving the radio emission. We neglect more detailed analysis such as considering spatial maps of the radio emission with comparison to observations of local deficits observed by \citet{murphy2009} as our CR model follows the protons whereas electrons generate the radio emission via synchrotron emission. Future work will use a spectral treatment of CR \citep[][]{yang2017spatially} to directly model the propagation and cooling of CR primary electrons, as well as the generation of secondaries. 

\subsection{Implications for Jellyfish Galaxies}
Recent observations have discovered multiphase tails of stripped gas, extending up to $\sim$100\,kpc from the disk (see \citealt{poggianti2017gaspSurvey}). Detections of molecular gas and star formation (e.g., \citealt{moretti2018gasp}) in these tails are difficult to reconcile with theoretical work suggesting dense gas is not readily stripped (e.g., \citealt{tonnesen2012star}). Regardless of the removal, cold, molecular gas should be rapidly heated and mixed with the ambient hot ICM (\citealt{Cowie1977}; \citealt{McKee1977}; \citealt{Balbus1982}; \citealt{Stone1992}; \citealt{Klein1994}; \citealt{Mac1994}; \citealt{Xu1995}). In contrast, the presence of star formation suggests stripped gas not only survives but attains sufficient mass to induce gravitational collapse.

High resolution 3D cloud-crushing simulations suggest efficient radiative cooling enable clouds to not only survive but grow in mass \citep[][Farber \& Gronke 2021]{Gronke2018,Gronke2020Cloudy,Sparre2020,Li2020,Kanjilal2021,Abruzzo2021}. \citet{Tonnesen2021} study the survival of cold clouds in RPS tails, finding the fate of cold gas may depend on ICM properties. Even in the destruction regime, molecular material may form on short time scales when dust is present \citep[][]{Girichidis2021}.

We find that CR provide nonthermal pressure support for a diffuse molecular phase (see Figure \ref{fig:PhasePlotNT}). At later times when a galaxy encounters the higher density of cluster cores, it is likely this diffuse molecular gas will be stripped, even if dense molecular cores remain unperturbed. If blobs of diffuse molecular gas are sufficiently large to satisfy the Farber \& Gronke criteria then these clouds may grow in mass and eventually become starforming. However, the impact of CR on cloud crushing remains relatively unexplored (see \citealt{wiener2017cosmic,Wiener2019,bruggen2020launching,Bustard2021}). In future work, we will explore these issues in greater depth.

\section{Caveats and Future Work}
\label{subsec:Caveats}
% Resolution of TS14: 159 pc
% Resolution of Ramos-Martinez+ 2017: 59 pc
% Resolution of R+14: 117 pc
\textit{Resolution.} Our physical resolution of $\sim$127\,pc is similar to that used in previous works including magnetic fields \citep{Tonnesen2014,Ruszkowski2014,ramos2018mhd}. Although we include nonthermal physics and stellar feedback neglected by previous hydrodynamical studies, it is true such hydro studies were performed at high resolution ($\sim$40\,pc) limiting the ability to compare with our work. We plan on running higher resolution models in future work.

\textit{Box Size.}
Our relatively small box size ($\sim$64 kpc)$^3$ and resolution degrading from the galactic disk limits our ability to study the fate of stripped gas: do CR enable stripping of diffuse molecular material that can survive and compose observed molecular tails? Do CR couple with the stripped gas? How do observed radio tails constrain CR transport? We plan to investigate such stimulating problems in future work.

\textit{Wind Profile.}
We utilized a relatively simple model for the relative motion between our simulated galaxy and ICM. That is, we assume a fixed magnetic field strength, density and temperature. We do vary the wind velocity, modeling the galaxy falling from rest into the cluster; however, we have a constant wind for $\sim$150\,Myr (for the ICM wind details, see \S \ref{subsec:NumericalTechniques}). In reality, a galaxy falling into a cluster should sample a hotter, denser, and faster wind as it plunges towards the center of the cluster and diminishing thereafter. We note that the time we simulate the galaxy only traverses $\lesssim$ 0.3\,Mpc. Thus, our galaxy samples cluster outskirts and our model is not implausible.

\ryChange{\textit{Momentum Feedback.} We utilized a momentum boost factor of 5 to account for unresolved clustering of supernovae as well as to account for advection errors. While observations clearly find the average kinetic energy ejected by supernovae to be $10^{51}$ erg, the coupling with the ambient ISM is highly uncertain. Several idealized, high-resolution studies have investigated the impact that clustered supernovae have on the momentum injected (coupled) to the ISM per supernova with varying results. \citet{kim2015momentum} find previous generations of bubbles decrease the injected momentum by a factor of two (similar to \citealt{pittard2019momentum}). Meanwhile, \citet{walch2015energy} find the momentum injected should be at least 25\% higher than the case of isolated supernovae, with \citet{gentry2017enhanced} finding up to an order of magnitude larger and \citet{keller2014superbubble} about five times larger. Additionally, \citet{diesing2018effect} find that cosmic rays may boost the momentum injected by up to an order of magnitude. Recently, \citet{montero2021momentum} performed 3D cosmic-ray magnetohydrodynamic simulations of isolated supernovae in homogeneous medium and found including cosmic rays increases the momentum deposition at least 50\% with a higher impact at lower ambient densities (as would be the case for clustered supernovae). Thus, our choice of a boost factor of five is within the plausible range, which will hopefully be more tightly constrained by future work. Moreover, since our results are based on comparisons between different runs with all runs using the same boost factor, we do not expect our results to be impacted.}

\textit{Neglected Physics.}
Cosmic rays are efficiently coupled to thermal plasma by scattering off waves they self-excite. The distribution of CR thus drifts with respect to the thermal plasma at the local Alfven velocity \citep[][]{kulsrud1969effect,zweibel2017basis}. However, wave damping processes such as ion-neutral friction \citep[][]{kulsrud1971effectiveness,Farber2018,Bustard2021}, turbulent damping \citep[][]{farmer2004wave,lazarian2016damping,holguin2019role}, linear Landau damping \citep[][]{wiener2018high} and nonlinear Landau damping \citep[][]{kulsrud2005plasma} enable CR to stream super-alfvenically and enable CR to heat the gas (mediated by the growth and subsequent damping of hydromagnetic waves). 

Our neglect of CR streaming in addition to collisional loss processes (e.g., Coulomb and hadronic) will reduce the CR energy, transferring it to the thermal gas. However, our \NoCR\ runs can effectively model CR losses proceeding very rapidly and thus comparison between the \NoCR, \ADV, and \DIF\ simulations should bracket the inclusion of CR losses. The enhanced transport of CR away from the cold, dense mid-plane due to ion-neutral damping may help to reduce CR suppression of the centralized accretion flow we observe in our simulations, see Figure \ref{fig:Accretion}. 

However, significant uncertainties in CR transport remain \citep[][]{hopkins2021testing}. Recent work suggests CR may not completely decouple in the neutral medium (due to the ``bottleneck'' effect; \citealt{wiener2017cosmic,Wiener2019,Bustard2021}) if pressure anisotropy can act as an efficient mechanism to grow hydromagnetic waves \citep{zweibel2020role}. Even in the absence of pressure anisotropy, dust grains may grow (or damp, depending on their transport relative to the Alven speed) hydromagnetic waves, even in molecular phases of low ionization fraction \citep[][]{Squire2021}. Exploring detailed models of (novel) cosmic-ray transport is beyond the scope of this work. Importantly, our results suggest CR transport can be effectively constrained in RPS studies, motivating future work on this topic.

We additionally ignore physics such as radiation pressure, anisotropic conduction, viscosity, and interactions with external galaxies as they are beyond the scope of this paper.

% ============================================================
% ============================================================
% ============================================================
% ============================================================
% ============================================================
% ============================================================
% ============================================================
% ============================================================
% ============================================================
% ============================================================
% ============================================================
% ============================================================
% ============================================================
% ============================================================
% ============================================================

\section{Conclusions}
\label{sec:Conclusions}
We performed the first cosmic-ray MHD simulations of an $L_{*}$ cluster galaxy subjected to ram pressure stripping, including radiative cooling, self-gravity of the gas, star formation, and stellar feedback. Our main conclusions are summarized as follows.

\begin{enumerate}
    \item Cosmic rays do not dramatically change the ram pressure stripping rate. This is true for the extreme cases of pure cosmic ray advection and diffusion without collisional losses and independent of the galaxy's inclination to the ICM wind. In all cases the stripping rates do not differ more than $\sim$15\% in the 230\,Myr of evolution we simulate.
    
    \item The nonthermal pressure provided by cosmic rays permits a stable low temperature, low density phase, which persists even with ram pressure stripping. Since low density gas is typically preferentially stripped, the additional cosmic ray pressure likely supports this gas against ram pressure stripping. Interestingly, ram pressure stripping increases the amount of low-temperature gas in disks surviving both face-on and edge-on winds, particularly when cosmic rays are included.
    
    \item The observed \textit{moderate} enhancement of star formation for cluster spirals undergoing ram pressure stripping places strong constraints on our models. Most of our simulations exhibit too large of a boost, attaining values $\sim$4-6 times greater for ram pressure stripped galaxies compared to isolated ones. We highlight that all of the \ADV\ simulations over-enhance SFRs with respect to observations, while the \FaceOn-\NoCR\ and \FaceOn-\DIF\ models are in plausible agreement with the observed enhancement.
    
    \item We find that cosmic rays suppress accretion of gas along the disk towards the midplane, even in isolated cases. Conversely, observations of cluster spirals indicating a high AGN fraction and bulge-dominated spirals (density-morphology relation) suggest enhanced gas accumulation in the nuclear region of galaxies undergoing ram pressure stripping. These observations appear to favor cosmic rays suffering rapid catastrophic losses or very efficient transport out of the disk. We suggest observations of ram pressure stripped galaxies may place novel constraints on cosmic-ray physics.
    
    \item In agreement with radio observations, our galaxy models suggest ram pressure stripped galaxies boast enhanced radio emission compared to their counterparts in the field. We find magnetic field strengths boosted by a factor of a few, and cosmic ray number densities enhanced by a factor of ten for the diffusion model and by a factor of a few for the advection model. Cosmic ray transport thus may play a crucial role in understanding the enhanced radio emission above the far-infrared to radio correlation for cluster spirals.
    
\end{enumerate}

%The Acknowledgements section is not numbered. Here you can thank helpful colleagues, acknowledge funding agencies, telescopes and facilities used etc. Try to keep it short.
\section*{Acknowledgements}

RJF gratefully acknowledges Dongwook Lee for helpful discussions.

RJF thanks the Kavli Institute for Theoretical Physics for hospitality during part of this project as part of the Graduate Fellowship Program. This research was supported in part by the National Science Foundation under Grant No. NSF PHY-1748958. % KITP
MR acknowledges support from NSF Collaborative Research Grants AST-1715140 and AST-2009227 and NASA grants 80NSSC20K1541 and 80NSSC20K1583.  ST acknowledges support from the Center for Computational Astrophysics at the Flatiron Institute, which is supported by
the Simons Foundation.
This project utilized the visualization and data analysis package \texttt{yt} \citep[][]{Turk2011}; we are grateful to the yt community for their support.

%%%%%%%%%%%%%%%%%%%%%%%%%%%%%%%%%%%%%%%%%%%%%%%%%%
\section*{Data Availability}

The simulation and data analysis scripts utilized to obtain the results presented in this paper will be shared on reasonable request to the corresponding author.



\bibliographystyle{mnras}
\bibliography{mnras_template}

\begin{thebibliography}{}
\makeatletter
\relax
\def\mn@urlcharsother{\let\do\@makeother \do\$\do\&\do\#\do\^\do\_\do\%\do\~}
\def\mn@doi{\begingroup\mn@urlcharsother \@ifnextchar [ {\mn@doi@}
  {\mn@doi@[]}}
\def\mn@doi@[#1]#2{\def\@tempa{#1}\ifx\@tempa\@empty \href
  {http://dx.doi.org/#2} {doi:#2}\else \href {http://dx.doi.org/#2} {#1}\fi
  \endgroup}
\def\mn@eprint#1#2{\mn@eprint@#1:#2::\@nil}
\def\mn@eprint@arXiv#1{\href {http://arxiv.org/abs/#1} {{\tt arXiv:#1}}}
\def\mn@eprint@dblp#1{\href {http://dblp.uni-trier.de/rec/bibtex/#1.xml}
  {dblp:#1}}
\def\mn@eprint@#1:#2:#3:#4\@nil{\def\@tempa {#1}\def\@tempb {#2}\def\@tempc
  {#3}\ifx \@tempc \@empty \let \@tempc \@tempb \let \@tempb \@tempa \fi \ifx
  \@tempb \@empty \def\@tempb {arXiv}\fi \@ifundefined
  {mn@eprint@\@tempb}{\@tempb:\@tempc}{\expandafter \expandafter \csname
  mn@eprint@\@tempb\endcsname \expandafter{\@tempc}}}

\bibitem[\protect\citeauthoryear{Abruzzo, Bryan  \& Fielding}{Abruzzo
  et~al.}{2021}]{Abruzzo2021}
Abruzzo M.~W.,  Bryan G.~L.,   Fielding D.~B.,  2021, arXiv preprint
  arXiv:2101.10344

\bibitem[\protect\citeauthoryear{Agertz, Kravtsov, Leitner  \& Gnedin}{Agertz
  et~al.}{2013}]{agertz2013toward}
Agertz O.,  Kravtsov A.~V.,  Leitner S.~N.,   Gnedin N.~Y.,  2013, The
  Astrophysical Journal, 770, 25

\bibitem[\protect\citeauthoryear{Balbus \& Hawley}{Balbus \&
  Hawley}{1991}]{balbus1991powerful}
Balbus S.~A.,  Hawley J.~F.,  1991, The Astrophysical Journal, 376, 214

\bibitem[\protect\citeauthoryear{Balbus \& McKee}{Balbus \&
  McKee}{1982}]{Balbus1982}
Balbus S.~A.,  McKee C.~F.,  1982, The Astrophysical Journal, 252, 529

\bibitem[\protect\citeauthoryear{Beck \& Golla}{Beck \& Golla}{1988}]{beck1988}
Beck R.,  Golla G.,  1988, Astronomy and Astrophysics, 191, L9

\bibitem[\protect\citeauthoryear{Bekki}{Bekki}{2013}]{bekki2013galactic}
Bekki K.,  2013, Monthly Notices of the Royal Astronomical Society, 438, 444

\bibitem[\protect\citeauthoryear{Booth, Agertz, Kravtsov  \& Gnedin}{Booth
  et~al.}{2013}]{booth2013simulations}
Booth C.,  Agertz O.,  Kravtsov A.~V.,   Gnedin N.~Y.,  2013, The Astrophysical
  Journal Letters, 777, L16

\bibitem[\protect\citeauthoryear{Boselli \& Gavazzi}{Boselli \&
  Gavazzi}{2006}]{boselli2006environmental}
Boselli A.,  Gavazzi G.,  2006, Publications of the Astronomical Society of the
  Pacific, 118, 517

\bibitem[\protect\citeauthoryear{Br{\"u}ggen \& Scannapieco}{Br{\"u}ggen \&
  Scannapieco}{2020}]{bruggen2020launching}
Br{\"u}ggen M.,  Scannapieco E.,  2020, The Astrophysical Journal, 905, 19

\bibitem[\protect\citeauthoryear{Burkert}{Burkert}{1995}]{burkert1995structure}
Burkert A.,  1995, The Astrophysical Journal Letters, 447, L25

\bibitem[\protect\citeauthoryear{Bustard \& Zweibel}{Bustard \&
  Zweibel}{2021}]{Bustard2021}
Bustard C.,  Zweibel E.~G.,  2021, The Astrophysical Journal, 913, 106

\bibitem[\protect\citeauthoryear{Bustard, Zweibel, D’Onghia, Gallagher~III
  \& Farber}{Bustard et~al.}{2020}]{bustard2020cosmic}
Bustard C.,  Zweibel E.~G.,  D’Onghia E.,  Gallagher~III J.,   Farber R.,
  2020, The Astrophysical Journal, 893, 29

\bibitem[\protect\citeauthoryear{Butcher \& Oemler~Jr}{Butcher \&
  Oemler~Jr}{1978}]{butcher1978b}
Butcher H.,  Oemler~Jr A.,  1978, The Astrophysical Journal, 226, 559

\bibitem[\protect\citeauthoryear{Butsky \& Quinn}{Butsky \&
  Quinn}{2018}]{butsky2018role}
Butsky I.~S.,  Quinn T.~R.,  2018, The Astrophysical Journal, 868, 108

\bibitem[\protect\citeauthoryear{Butsky, Fielding, Hayward, Hummels, Quinn  \&
  Werk}{Butsky et~al.}{2020}]{Butsky2020}
Butsky I.~S.,  Fielding D.~B.,  Hayward C.~C.,  Hummels C.~B.,  Quinn T.~R.,
  Werk J.~K.,  2020, The Astrophysical Journal, 903, 77

\bibitem[\protect\citeauthoryear{Chan, Kereš, Hopkins, Quataert, Su, Hayward,
  Faucher  \& Ere}{Chan et~al.}{2019}]{Chan2019}
Chan T.~K.,  Kereš D.,  Hopkins P.~F.,  Quataert E.,  Su K.-Y.,  Hayward
  C.~C.,  Faucher C.-A.,   Ere G.,  2019, \mn@doi [MNRAS]
  {10.1093/mnras/stz1895}, 488, 3716

\bibitem[\protect\citeauthoryear{Cowie \& McKee}{Cowie \&
  McKee}{1977}]{Cowie1977}
Cowie L.~L.,  McKee C.~F.,  1977, The Astrophysical Journal, 211, 135

\bibitem[\protect\citeauthoryear{Dickey \& Salpeter}{Dickey \&
  Salpeter}{1984}]{dickey19841}
Dickey J.,  Salpeter E.,  1984, The Astrophysical Journal, 284, 461

\bibitem[\protect\citeauthoryear{Diesing \& Caprioli}{Diesing \&
  Caprioli}{2018}]{diesing2018effect}
Diesing R.,  Caprioli D.,  2018, Physical review letters, 121, 091101

\bibitem[\protect\citeauthoryear{Dressler}{Dressler}{1980}]{dressler1980galaxy}
Dressler A.,  1980, The Astrophysical Journal, 236, 351

\bibitem[\protect\citeauthoryear{Dubey, Reid  \& Fisher}{Dubey
  et~al.}{2008}]{Dubey2008}
Dubey A.,  Reid L.~B.,   Fisher R.,  2008, Physica Scripta, 2008, 14046

\bibitem[\protect\citeauthoryear{Dursi \& Pfrommer}{Dursi \&
  Pfrommer}{2008}]{dursi2008draping}
Dursi L.,  Pfrommer C.,  2008, The Astrophysical Journal, 677, 993

\bibitem[\protect\citeauthoryear{En{\ss}lin, Pfrommer, Springel  \&
  Jubelgas}{En{\ss}lin et~al.}{2007}]{ensslin2007cosmic}
En{\ss}lin T.~A.,  Pfrommer C.,  Springel V.,   Jubelgas M.,  2007, Astronomy
  \& Astrophysics, 473, 41

\bibitem[\protect\citeauthoryear{Everett, Zweibel, Benjamin, McCammon, Rocks
  \& Gallagher~III}{Everett et~al.}{2008}]{everett2008milky}
Everett J.~E.,  Zweibel E.~G.,  Benjamin R.~A.,  McCammon D.,  Rocks L.,
  Gallagher~III J.~S.,  2008, The Astrophysical Journal, 674, 258

\bibitem[\protect\citeauthoryear{Farber, Ruszkowski, Yang  \& Zweibel}{Farber
  et~al.}{2018}]{Farber2018}
Farber R.,  Ruszkowski M.,  Yang H.-Y.,   Zweibel E.~G.,  2018, The
  Astrophysical Journal, 856, 112

\bibitem[\protect\citeauthoryear{Farmer \& Goldreich}{Farmer \&
  Goldreich}{2004}]{farmer2004wave}
Farmer A.~J.,  Goldreich P.,  2004, The Astrophysical Journal, 604, 671

\bibitem[\protect\citeauthoryear{Ferland, Korista, Verner, Ferguson, Kingdon
  \& Verner}{Ferland et~al.}{1998}]{ferland1998cloudy}
Ferland G.,  Korista K.,  Verner D.,  Ferguson J.,  Kingdon J.,   Verner E.,
  1998, Publications of the Astronomical Society of the Pacific, 110, 761

\bibitem[\protect\citeauthoryear{Fryxell et~al.,}{Fryxell
  et~al.}{2000}]{Fryxell2000}
Fryxell B.,  et~al., 2000, \mn@doi [The Astrophysical Journal Supplement
  Series] {10.1086/317361}, 131, 273

\bibitem[\protect\citeauthoryear{Gentry, Krumholz, Dekel  \& Madau}{Gentry
  et~al.}{2017}]{gentry2017enhanced}
Gentry E.~S.,  Krumholz M.~R.,  Dekel A.,   Madau P.,  2017, Monthly Notices of
  the Royal Astronomical Society, 465, 2471

\bibitem[\protect\citeauthoryear{Gentry, Krumholz, Madau  \& Lupi}{Gentry
  et~al.}{2019}]{gentry2019momentum}
Gentry E.~S.,  Krumholz M.~R.,  Madau P.,   Lupi A.,  2019, Monthly Notices of
  the Royal Astronomical Society, 483, 3647

\bibitem[\protect\citeauthoryear{Girichidis et~al.,}{Girichidis
  et~al.}{2016}]{girichidis2016launching}
Girichidis P.,  et~al., 2016, The Astrophysical Journal Letters, 816, L19

\bibitem[\protect\citeauthoryear{Girichidis, Naab, Hanasz  \& Walch}{Girichidis
  et~al.}{2018}]{girichidis2018cooler}
Girichidis P.,  Naab T.,  Hanasz M.,   Walch S.,  2018, Monthly Notices of the
  Royal Astronomical Society, 479, 3042

\bibitem[\protect\citeauthoryear{Girichidis, Naab, Walch  \& Berlok}{Girichidis
  et~al.}{2021}]{Girichidis2021}
Girichidis P.,  Naab T.,  Walch S.,   Berlok T.,  2021, arXiv preprint
  arXiv:2101.08269

\bibitem[\protect\citeauthoryear{Gnedin \& Kravtsov}{Gnedin \&
  Kravtsov}{2011}]{gnedin2011environmental}
Gnedin N.~Y.,  Kravtsov A.~V.,  2011, The Astrophysical Journal, 728, 88

\bibitem[\protect\citeauthoryear{Grenier, Black  \& Strong}{Grenier
  et~al.}{2015}]{grenier2015nine}
Grenier I.~A.,  Black J.~H.,   Strong A.~W.,  2015, Annual Review of Astronomy
  and Astrophysics, 53, 199

\bibitem[\protect\citeauthoryear{Gronke \& Oh}{Gronke \& Oh}{2018}]{Gronke2018}
Gronke M.,  Oh S.~P.,  2018, Monthly Notices of the Royal Astronomical Society:
  Letters, 480, L111–L115

\bibitem[\protect\citeauthoryear{Gronke \& Oh}{Gronke \&
  Oh}{2020}]{Gronke2020Cloudy}
Gronke M.,  Oh S.~P.,  2020, Monthly Notices of the Royal Astronomical Society,
  492, 1970

\bibitem[\protect\citeauthoryear{Gunn \& Gott~III}{Gunn \&
  Gott~III}{1972}]{gunn1972infall}
Gunn J.~E.,  Gott~III J.~R.,  1972, The Astrophysical Journal, 176, 1

\bibitem[\protect\citeauthoryear{Hawley \& Balbus}{Hawley \&
  Balbus}{1999}]{hawley1999transport}
Hawley J.~F.,  Balbus S.~A.,  1999, Physics of Plasmas, 6, 4444

\bibitem[\protect\citeauthoryear{Heintz \& Zweibel}{Heintz \&
  Zweibel}{2018}]{heintz2018parker}
Heintz E.,  Zweibel E.~G.,  2018, The Astrophysical Journal, 860, 97

\bibitem[\protect\citeauthoryear{Hernquist}{Hernquist}{1993}]{hernquist1993n}
Hernquist L.,  1993, The Astrophysical Journal Supplement Series, 86, 389

\bibitem[\protect\citeauthoryear{Holguin, Ruszkowski, Lazarian, Farber  \&
  Yang}{Holguin et~al.}{2019}]{holguin2019role}
Holguin F.,  Ruszkowski M.,  Lazarian A.,  Farber R.,   Yang H.~K.,  2019,
  Monthly Notices of the Royal Astronomical Society, 490, 1271

\bibitem[\protect\citeauthoryear{Hopkins et~al.,}{Hopkins
  et~al.}{2020}]{hopkins2020but}
Hopkins P.~F.,  et~al., 2020, Monthly Notices of the Royal Astronomical
  Society, 492, 3465

\bibitem[\protect\citeauthoryear{Hopkins, Squire, Chan, Quataert, Ji,
  Kere{\v{s}}  \& Faucher-Gigu{\`e}re}{Hopkins
  et~al.}{2021}]{hopkins2021testing}
Hopkins P.~F.,  Squire J.,  Chan T.,  Quataert E.,  Ji S.,  Kere{\v{s}} D.,
  Faucher-Gigu{\`e}re C.-A.,  2021, Monthly Notices of the Royal Astronomical
  Society, 501, 4184

\bibitem[\protect\citeauthoryear{Huang \& Greengard}{Huang \&
  Greengard}{1999}]{huang1999fast}
Huang J.,  Greengard L.,  1999, SIAM Journal on Scientific Computing, 21, 1551

\bibitem[\protect\citeauthoryear{Hubble \& Humason}{Hubble \&
  Humason}{1931}]{hubble1931velocity}
Hubble E.,  Humason M.~L.,  1931, The Astrophysical Journal, 74, 43

\bibitem[\protect\citeauthoryear{Iffrig \& Hennebelle}{Iffrig \&
  Hennebelle}{2015}]{iffrig2015mutual}
Iffrig O.,  Hennebelle P.,  2015, Astronomy \& Astrophysics, 576, A95

\bibitem[\protect\citeauthoryear{J{\'a}chym, K{\"o}ppen, Palou{\v{s}}  \&
  Combes}{J{\'a}chym et~al.}{2009}]{jachym2009ram}
J{\'a}chym P.,  K{\"o}ppen J.,  Palou{\v{s}} J.,   Combes F.,  2009, Astronomy
  \& Astrophysics, 500, 693

\bibitem[\protect\citeauthoryear{J{\'a}chym et~al.,}{J{\'a}chym
  et~al.}{2017}]{jachym2017molecular}
J{\'a}chym P.,  et~al., 2017, The Astrophysical Journal, 839, 114

\bibitem[\protect\citeauthoryear{Jaff{\'e}, Smith, Candlish, Poggianti, Sheen
  \& Verheijen}{Jaff{\'e} et~al.}{2015}]{jaffe2015budhies}
Jaff{\'e} Y.~L.,  Smith R.,  Candlish G.~N.,  Poggianti B.~M.,  Sheen Y.-K.,
  Verheijen M.~A.,  2015, Monthly Notices of the Royal Astronomical Society,
  448, 1715

\bibitem[\protect\citeauthoryear{Ji et~al.,}{Ji
  et~al.}{2020}]{ji2020properties}
Ji S.,  et~al., 2020, Monthly Notices of the Royal Astronomical Society, 496,
  4221

\bibitem[\protect\citeauthoryear{Kanjilal, Dutta  \& Sharma}{Kanjilal
  et~al.}{2021}]{Kanjilal2021}
Kanjilal V.,  Dutta A.,   Sharma P.,  2021, Monthly Notices of the Royal
  Astronomical Society, 501, 1143

\bibitem[\protect\citeauthoryear{Keller, Wadsley, Benincasa  \&
  Couchman}{Keller et~al.}{2014}]{keller2014superbubble}
Keller B.,  Wadsley J.,  Benincasa S.,   Couchman H.,  2014, Monthly Notices of
  the Royal Astronomical Society, 442, 3013

\bibitem[\protect\citeauthoryear{Kennicutt~Jr}{Kennicutt~Jr}{1998}]{kennicutt1998global}
Kennicutt~Jr R.~C.,  1998, The Astrophysical Journal, 498, 541

\bibitem[\protect\citeauthoryear{Kim \& Ostriker}{Kim \&
  Ostriker}{2015}]{kim2015momentum}
Kim C.-G.,  Ostriker E.~C.,  2015, The Astrophysical Journal, 802, 99

\bibitem[\protect\citeauthoryear{Kim \& Ostriker}{Kim \&
  Ostriker}{2017}]{kim2017three}
Kim C.-G.,  Ostriker E.~C.,  2017, The Astrophysical Journal, 846, 133

\bibitem[\protect\citeauthoryear{Klein, McKee  \& Colella}{Klein
  et~al.}{1994}]{Klein1994}
Klein R.~I.,  McKee C.~F.,   Colella P.,  1994, The Astrophysical Journal, 420,
  213

\bibitem[\protect\citeauthoryear{Kravtsov}{Kravtsov}{1999}]{kravtsov1999phd}
Kravtsov A.,  1999, PhD thesis, New Mexico State Univ.

\bibitem[\protect\citeauthoryear{Kravtsov, Klypin  \& Hoffman}{Kravtsov
  et~al.}{2002}]{kravtsov2002constrained}
Kravtsov A.~V.,  Klypin A.,   Hoffman Y.,  2002, The Astrophysical Journal,
  571, 563

\bibitem[\protect\citeauthoryear{Kronberger, Kapferer, Ferrari,
  Unterguggenberger  \& Schindler}{Kronberger
  et~al.}{2008}]{kronberger2008influence}
Kronberger T.,  Kapferer W.,  Ferrari C.,  Unterguggenberger S.,   Schindler
  S.,  2008, Astronomy \& Astrophysics, 481, 337

\bibitem[\protect\citeauthoryear{Krumholz \& Tan}{Krumholz \&
  Tan}{2007}]{krumholz2007slow}
Krumholz M.~R.,  Tan J.~C.,  2007, The Astrophysical Journal, 654, 304

\bibitem[\protect\citeauthoryear{Kulsrud}{Kulsrud}{2005}]{kulsrud2005plasma}
Kulsrud R.~M.,  2005, RM Kulsrud

\bibitem[\protect\citeauthoryear{Kulsrud \& Cesarsky}{Kulsrud \&
  Cesarsky}{1971}]{kulsrud1971effectiveness}
Kulsrud R.,  Cesarsky C.,  1971, Astrophysical Letters, 8, 189

\bibitem[\protect\citeauthoryear{Kulsrud \& Pearce}{Kulsrud \&
  Pearce}{1969}]{kulsrud1969effect}
Kulsrud R.,  Pearce W.~P.,  1969, The Astrophysical Journal, 156, 445

\bibitem[\protect\citeauthoryear{Lacki, Thompson  \& Quataert}{Lacki
  et~al.}{2010}]{lacki2010}
Lacki B.~C.,  Thompson T.~A.,   Quataert E.,  2010, The Astrophysical Journal,
  717, 1

\bibitem[\protect\citeauthoryear{Lazarian}{Lazarian}{2016}]{lazarian2016damping}
Lazarian A.,  2016, The Astrophysical Journal, 833, 131

\bibitem[\protect\citeauthoryear{Lee}{Lee}{2013}]{Lee2013}
Lee D.,  2013, \mn@doi [Journal of Computational Physics]
  {10.1016/j.jcp.2013.02.049}, 243, 269

\bibitem[\protect\citeauthoryear{Lee \& Deane}{Lee \& Deane}{2009}]{Lee2009}
Lee D.,  Deane A.~E.,  2009, \mn@doi [Journal of Computational Physics]
  {10.1016/j.jcp.2008.08.026}, 228

\bibitem[\protect\citeauthoryear{Lee, Kimm, Katz, Rosdahl, Devriendt  \&
  Slyz}{Lee et~al.}{2020}]{lee2020dual}
Lee J.,  Kimm T.,  Katz H.,  Rosdahl J.,  Devriendt J.,   Slyz A.,  2020, The
  Astrophysical Journal, 905, 31

\bibitem[\protect\citeauthoryear{Leroy et~al.,}{Leroy
  et~al.}{2017}]{leroy2017cloud}
Leroy A.~K.,  et~al., 2017, The Astrophysical Journal, 846, 71

\bibitem[\protect\citeauthoryear{Li, Hopkins, Squire  \& Hummels}{Li
  et~al.}{2020}]{Li2020}
Li Z.,  Hopkins P.~F.,  Squire J.,   Hummels C.,  2020, \mn@doi [Monthly
  Notices of the Royal Astronomical Society] {10.1093/mnras/stz3567}, 492, 1841

\bibitem[\protect\citeauthoryear{Lyutikov}{Lyutikov}{2006}]{lyutikov2006magnetic}
Lyutikov M.,  2006, Monthly Notices of the Royal Astronomical Society, 373, 73

\bibitem[\protect\citeauthoryear{{Mac Low}, McKee, Klein, Stone  \&
  Norman}{{Mac Low} et~al.}{1994}]{Mac1994}
{Mac Low} M.-M.,  McKee C.~F.,  Klein R.~I.,  Stone J.~M.,   Norman M.~L.,
  1994, The Astrophysical Journal, 433, 757

\bibitem[\protect\citeauthoryear{Martizzi, Faucher-Gigu{\`e}re  \&
  Quataert}{Martizzi et~al.}{2015}]{martizzi2015supernova}
Martizzi D.,  Faucher-Gigu{\`e}re C.-A.,   Quataert E.,  2015, Monthly Notices
  of the Royal Astronomical Society, 450, 504

\bibitem[\protect\citeauthoryear{McKee \& Cowie}{McKee \&
  Cowie}{1977}]{McKee1977}
McKee C.~F.,  Cowie L.~L.,  1977, The Astrophysical Journal, 215, 213

\bibitem[\protect\citeauthoryear{Miyamoto \& Nagai}{Miyamoto \&
  Nagai}{1975}]{miyamoto1975three}
Miyamoto M.,  Nagai R.,  1975, Publications of the Astronomical Society of
  Japan, 27, 533

\bibitem[\protect\citeauthoryear{Montero, Martin-Alvarez, Sijacki, Slyz,
  Devriendt  \& Dubois}{Montero et~al.}{2021}]{montero2021momentum}
Montero F.~R.,  Martin-Alvarez S.,  Sijacki D.,  Slyz A.,  Devriendt J.,
  Dubois Y.,  2021, arXiv preprint arXiv:2110.09862

\bibitem[\protect\citeauthoryear{Moretti et~al.,}{Moretti
  et~al.}{2018}]{moretti2018gasp}
Moretti A.,  et~al., 2018, Monthly Notices of the Royal Astronomical Society,
  480, 2508

\bibitem[\protect\citeauthoryear{Moretti et~al.,}{Moretti
  et~al.}{2020a}]{moretti2020gasp}
Moretti A.,  et~al., 2020a, The Astrophysical Journal, 889, 9

\bibitem[\protect\citeauthoryear{Moretti et~al.,}{Moretti
  et~al.}{2020b}]{moretti2020high}
Moretti A.,  et~al., 2020b, The Astrophysical Journal Letters, 897, L30

\bibitem[\protect\citeauthoryear{M{\"u}ller et~al.,}{M{\"u}ller
  et~al.}{2021}]{muller2021highly}
M{\"u}ller A.,  et~al., 2021, Nature Astronomy, 5, 159

\bibitem[\protect\citeauthoryear{Murphy, Kenney, Helou, Chung  \&
  Howell}{Murphy et~al.}{2009}]{murphy2009}
Murphy E.~J.,  Kenney J. D.~P.,  Helou G.,  Chung A.,   Howell J.~H.,  2009,
  The Astrophysical Journal, 694, 1435

\bibitem[\protect\citeauthoryear{Oosterloo \& van Gorkom}{Oosterloo \& van
  Gorkom}{2005}]{oosterloo2005large}
Oosterloo T.,  van Gorkom J.,  2005, Astronomy \& Astrophysics, 437, L19

\bibitem[\protect\citeauthoryear{Paladino, Murgia, Helfer, Wong, Ekers, Blitz,
  Gregorini  \& Moscadelli}{Paladino et~al.}{2006}]{paladino2006}
Paladino R.,  Murgia M.,  Helfer T.~T.,  Wong T.,  Ekers R.,  Blitz L.,
  Gregorini L.,   Moscadelli L.,  2006, Astronomy and Astrophysics, 456, 847

\bibitem[\protect\citeauthoryear{Pfrommer \& Dursi}{Pfrommer \&
  Dursi}{2010}]{pfrommer2010detecting}
Pfrommer C.,  Dursi J.,  2010, Nature Physics, 6, 520

\bibitem[\protect\citeauthoryear{Pfrommer, Pakmor, Schaal, Simpson  \&
  Springel}{Pfrommer et~al.}{2017a}]{pfrommer2017simulating}
Pfrommer C.,  Pakmor R.,  Schaal K.,  Simpson C.,   Springel V.,  2017a,
  Monthly Notices of the Royal Astronomical Society, 465, 4500

\bibitem[\protect\citeauthoryear{Pfrommer, Pakmor, Simpson  \&
  Springel}{Pfrommer et~al.}{2017b}]{pfrommer2017gamma}
Pfrommer C.,  Pakmor R.,  Simpson C.~M.,   Springel V.,  2017b, The
  Astrophysical Journal Letters, 847, L13

\bibitem[\protect\citeauthoryear{Pittard}{Pittard}{2019}]{pittard2019momentum}
Pittard J.~M.,  2019, Monthly Notices of the Royal Astronomical Society, 488,
  3376

\bibitem[\protect\citeauthoryear{Poggianti et~al.,}{Poggianti
  et~al.}{2017a}]{poggianti2017agn}
Poggianti B.~M.,  et~al., 2017a, Nature, 548, 304

\bibitem[\protect\citeauthoryear{Poggianti et~al.,}{Poggianti
  et~al.}{2017b}]{poggianti2017gaspSurvey}
Poggianti B.~M.,  et~al., 2017b, The Astrophysical Journal, 844, 48

\bibitem[\protect\citeauthoryear{Radovich, Poggianti, Jaff{\'e}, Moretti,
  Bettoni, Gullieuszik, Vulcani  \& Fritz}{Radovich
  et~al.}{2019}]{radovich2019gasp}
Radovich M.,  Poggianti B.,  Jaff{\'e} Y.~L.,  Moretti A.,  Bettoni D.,
  Gullieuszik M.,  Vulcani B.,   Fritz J.,  2019, Monthly Notices of the Royal
  Astronomical Society, 486, 486

\bibitem[\protect\citeauthoryear{Ramos-Mart{\'\i}nez, G{\'o}mez  \&
  P{\'e}rez-Villegas}{Ramos-Mart{\'\i}nez et~al.}{2018}]{ramos2018mhd}
Ramos-Mart{\'\i}nez M.,  G{\'o}mez G.~C.,   P{\'e}rez-Villegas {\'A}.,  2018,
  Monthly Notices of the Royal Astronomical Society, 476, 3781

\bibitem[\protect\citeauthoryear{Ricker}{Ricker}{2008}]{ricker2008direct}
Ricker P.,  2008, The Astrophysical Journal Supplement Series, 176, 293

\bibitem[\protect\citeauthoryear{Roediger}{Roediger}{2009}]{roediger2009ram}
Roediger E.,  2009, Astronomische Nachrichten: Astronomical Notes, 330, 888

\bibitem[\protect\citeauthoryear{Roediger \& Br{\"u}ggen}{Roediger \&
  Br{\"u}ggen}{2006}]{roediger2006ram}
Roediger E.,  Br{\"u}ggen M.,  2006, Monthly Notices of the Royal Astronomical
  Society, 369, 567

\bibitem[\protect\citeauthoryear{Roediger, Br{\"u}ggen, Owers, Ebeling  \&
  Sun}{Roediger et~al.}{2014}]{roediger2014star}
Roediger E.,  Br{\"u}ggen M.,  Owers M.,  Ebeling H.,   Sun M.,  2014, Monthly
  Notices of the Royal Astronomical Society: Letters, 443, L114

\bibitem[\protect\citeauthoryear{Rudd, Zentner  \& Kravtsov}{Rudd
  et~al.}{2008}]{rudd2008effects}
Rudd D.~H.,  Zentner A.~R.,   Kravtsov A.~V.,  2008, The Astrophysical Journal,
  672, 19

\bibitem[\protect\citeauthoryear{Ruggiero \& Lima~Neto}{Ruggiero \&
  Lima~Neto}{2017}]{ruggiero2017fate}
Ruggiero R.,  Lima~Neto G.~B.,  2017, Monthly Notices of the Royal Astronomical
  Society, 468, 4107

\bibitem[\protect\citeauthoryear{Ruszkowski, En{\ss}lin, Br{\"u}ggen, Heinz  \&
  Pfrommer}{Ruszkowski et~al.}{2007}]{ruszkowski2007impact}
Ruszkowski M.,  En{\ss}lin T.,  Br{\"u}ggen M.,  Heinz S.,   Pfrommer C.,
  2007, Monthly Notices of the Royal Astronomical Society, 378, 662

\bibitem[\protect\citeauthoryear{Ruszkowski, En{\ss}lin, Br{\"u}ggen, Begelman
  \& Churazov}{Ruszkowski et~al.}{2008}]{ruszkowski2008cosmic}
Ruszkowski M.,  En{\ss}lin T.,  Br{\"u}ggen M.,  Begelman M.,   Churazov E.,
  2008, Monthly Notices of the Royal Astronomical Society, 383, 1359

\bibitem[\protect\citeauthoryear{Ruszkowski, Brüggen, Lee  \& Shin}{Ruszkowski
  et~al.}{2014}]{Ruszkowski2014}
Ruszkowski M.,  Brüggen M.,  Lee D.,   Shin M.-S.,  2014, \mn@doi [The
  Astrophysical Journal] {10.1088/0004-637X/784/1/75}, 784, 75

\bibitem[\protect\citeauthoryear{Ruszkowski, Yang  \& Zweibel}{Ruszkowski
  et~al.}{2017}]{ruszkowski2017global}
Ruszkowski M.,  Yang H.-Y.~K.,   Zweibel E.,  2017, The Astrophysical Journal,
  834, 208

\bibitem[\protect\citeauthoryear{Salem \& Bryan}{Salem \&
  Bryan}{2014}]{salem2014cosmic}
Salem M.,  Bryan G.~L.,  2014, Monthly Notices of the Royal Astronomical
  Society, 437, 3312

\bibitem[\protect\citeauthoryear{Schmidt}{Schmidt}{1959}]{schmidt1959rate}
Schmidt M.,  1959, The astrophysical journal, 129, 243

\bibitem[\protect\citeauthoryear{Schulz \& Struck}{Schulz \&
  Struck}{2001}]{schulz2001multi}
Schulz S.,  Struck C.,  2001, Monthly Notices of the Royal Astronomical
  Society, 328, 185

\bibitem[\protect\citeauthoryear{Scodeggio \& Gavazzi}{Scodeggio \&
  Gavazzi}{1993}]{scodeggio1993}
Scodeggio M.,  Gavazzi G.,  1993, The Astrophysical Journal, 409, 110

\bibitem[\protect\citeauthoryear{Semenov, Kravtsov  \& Gnedin}{Semenov
  et~al.}{2016}]{semenov2016nonuniversal}
Semenov V.~A.,  Kravtsov A.~V.,   Gnedin N.~Y.,  2016, The Astrophysical
  Journal, 826, 200

\bibitem[\protect\citeauthoryear{Semenov, Kravtsov  \& Gnedin}{Semenov
  et~al.}{2017}]{semenov2017physical}
Semenov V.~A.,  Kravtsov A.~V.,   Gnedin N.~Y.,  2017, The Astrophysical
  Journal, 845, 133

\bibitem[\protect\citeauthoryear{Semenov, Kravtsov  \& Gnedin}{Semenov
  et~al.}{2018}]{semenov2018galaxies}
Semenov V.~A.,  Kravtsov A.~V.,   Gnedin N.~Y.,  2018, The Astrophysical
  Journal, 861, 4

\bibitem[\protect\citeauthoryear{Semenov, Kravtsov  \& Caprioli}{Semenov
  et~al.}{2021}]{semenov2021cosmic}
Semenov V.~A.,  Kravtsov A.~V.,   Caprioli D.,  2021, The Astrophysical
  Journal, 910, 126

\bibitem[\protect\citeauthoryear{Simpson, Pakmor, Marinacci, Pfrommer,
  Springel, Glover, Clark  \& Smith}{Simpson et~al.}{2016}]{simpson2016role}
Simpson C.~M.,  Pakmor R.,  Marinacci F.,  Pfrommer C.,  Springel V.,  Glover
  S.~C.,  Clark P.~C.,   Smith R.~J.,  2016, The Astrophysical Journal Letters,
  827, L29

\bibitem[\protect\citeauthoryear{Sparre, Pfrommer  \& Ehlert}{Sparre
  et~al.}{2020}]{Sparre2020}
Sparre M.,  Pfrommer C.,   Ehlert K.,  2020, \mn@doi [Monthly Notices of the
  Royal Astronomical Society] {10.1093/mnras/staa3177}, 499, 4261

\bibitem[\protect\citeauthoryear{Squire, Hopkins, Quataert  \& Kempski}{Squire
  et~al.}{2021}]{Squire2021}
Squire J.,  Hopkins P.~F.,  Quataert E.,   Kempski P.,  2021, \mn@doi [MNRAS]
  {10.1093/mnras/stab179}, 502, 2630

\bibitem[\protect\citeauthoryear{Steinhauser, Schindler  \&
  Springel}{Steinhauser et~al.}{2016}]{steinhauser2016simulations}
Steinhauser D.,  Schindler S.,   Springel V.,  2016, Astronomy \& Astrophysics,
  591, A51

\bibitem[\protect\citeauthoryear{Stone \& Norman}{Stone \&
  Norman}{1992}]{Stone1992}
Stone J.~M.,  Norman M.~L.,  1992, The Astrophysical Journal, 173, 17

\bibitem[\protect\citeauthoryear{Su, Slatyer  \& Finkbeiner}{Su
  et~al.}{2010}]{su2010giant}
Su M.,  Slatyer T.~R.,   Finkbeiner D.~P.,  2010, The Astrophysical Journal,
  724, 1044

\bibitem[\protect\citeauthoryear{Sun, Jones, Forman, Nulsen, Donahue  \&
  Voit}{Sun et~al.}{2006}]{sun200670}
Sun M.,  Jones C.,  Forman W.,  Nulsen P.,  Donahue M.,   Voit G.,  2006, The
  Astrophysical Journal Letters, 637, L81

\bibitem[\protect\citeauthoryear{Sun, Donahue, Roediger, Nulsen, Voit, Sarazin,
  Forman  \& Jones}{Sun et~al.}{2009}]{sun2009spectacular}
Sun M.,  Donahue M.,  Roediger E.,  Nulsen P.,  Voit G.,  Sarazin C.,  Forman
  W.,   Jones C.,  2009, The Astrophysical Journal, 708, 946

\bibitem[\protect\citeauthoryear{Tonnesen}{Tonnesen}{2019}]{Tonnesen2019}
Tonnesen S.,  2019, \mn@doi [The Astrophysical Journal]
  {10.3847/1538-4357/ab0960}, 874, 161

\bibitem[\protect\citeauthoryear{Tonnesen \& Bryan}{Tonnesen \&
  Bryan}{2009}]{tonnesen2009gas}
Tonnesen S.,  Bryan G.~L.,  2009, The Astrophysical Journal, 694, 789

\bibitem[\protect\citeauthoryear{Tonnesen \& Bryan}{Tonnesen \&
  Bryan}{2010}]{tonnesen2010tail}
Tonnesen S.,  Bryan G.~L.,  2010, The Astrophysical Journal, 709, 1203

\bibitem[\protect\citeauthoryear{Tonnesen \& Bryan}{Tonnesen \&
  Bryan}{2012}]{tonnesen2012star}
Tonnesen S.,  Bryan G.~L.,  2012, Monthly Notices of the Royal Astronomical
  Society, 422, 1609

\bibitem[\protect\citeauthoryear{Tonnesen \& Bryan}{Tonnesen \&
  Bryan}{2021}]{Tonnesen2021}
Tonnesen S.,  Bryan G.~L.,  2021, The Astrophysical Journal, 911, 68

\bibitem[\protect\citeauthoryear{Tonnesen \& Stone}{Tonnesen \&
  Stone}{2014}]{Tonnesen2014}
Tonnesen S.,  Stone J.,  2014, \mn@doi [The Astrophysical Journal]
  {10.1088/0004-637X/795/2/148}, 795, 148

\bibitem[\protect\citeauthoryear{Trachternach, De~Blok, Walter, Brinks  \&
  Kennicutt~Jr}{Trachternach et~al.}{2008}]{trachternach2008dynamical}
Trachternach C.,  De~Blok W.,  Walter F.,  Brinks E.,   Kennicutt~Jr R.,  2008,
  The Astronomical Journal, 136, 2720

\bibitem[\protect\citeauthoryear{Trapp et~al.,}{Trapp
  et~al.}{2021}]{trapp2021gas}
Trapp C.,  et~al., 2021, arXiv preprint arXiv:2105.11472

\bibitem[\protect\citeauthoryear{Turk, Smith, Oishi, Skory, Skillman, Abel  \&
  Norman}{Turk et~al.}{2010}]{Turk2011}
Turk M.~J.,  Smith B.~D.,  Oishi J.~S.,  Skory S.,  Skillman S.~W.,  Abel T.,
  Norman M.~L.,  2010, \apjs, 192, 9

\bibitem[\protect\citeauthoryear{Uhlig, Pfrommer, Sharma, Nath, En{\ss}lin  \&
  Springel}{Uhlig et~al.}{2012}]{uhlig2012galactic}
Uhlig M.,  Pfrommer C.,  Sharma M.,  Nath B.~B.,  En{\ss}lin T.,   Springel V.,
   2012, Monthly Notices of the Royal Astronomical Society, 423, 2374

\bibitem[\protect\citeauthoryear{Vollmer, Soida, Chung, Chemin, Braine, Boselli
   \& Beck}{Vollmer et~al.}{2009}]{vollmer2009}
Vollmer B.,  Soida M.,  Chung A.,  Chemin L.,  Braine J.,  Boselli A.,   Beck
  R.,  2009, Astronomy and Astrophysics, 496, 669

\bibitem[\protect\citeauthoryear{Vollmer, Soida, Chung, Beck, Urbanik, Chyży,
  Otmianowska-Mazur  \& Gorkom}{Vollmer et~al.}{2010}]{vollmer2010}
Vollmer B.,  Soida M.,  Chung A.,  Beck R.,  Urbanik M.,  Chyży K.~T.,
  Otmianowska-Mazur K.,   Gorkom J. H.~V.,  2010, Astronomy and Astrophysics,
  512, A36

\bibitem[\protect\citeauthoryear{Vollmer, Soida, Beck, Chung, Urbanik, Chyży,
  Otmianowska-Mazur  \& Kenney}{Vollmer et~al.}{2013}]{vollmer2013}
Vollmer B.,  Soida M.,  Beck R.,  Chung A.,  Urbanik M.,  Chyży K.~T.,
  Otmianowska-Mazur K.,   Kenney J. D.~P.,  2013, Astronomy and Astrophysics,
  553, A116

\bibitem[\protect\citeauthoryear{Vulcani et~al.,}{Vulcani
  et~al.}{2018}]{vulcani2018enhanced}
Vulcani B.,  et~al., 2018, The Astrophysical Journal Letters, 866, L25

\bibitem[\protect\citeauthoryear{Walch \& Naab}{Walch \&
  Naab}{2015}]{walch2015energy}
Walch S.,  Naab T.,  2015, Monthly Notices of the Royal Astronomical Society,
  451, 2757

\bibitem[\protect\citeauthoryear{Werk, Prochaska, Thom, Tumlinson, Tripp,
  O'Meara  \& Peeples}{Werk et~al.}{2013}]{Werk2013}
Werk J.~K.,  Prochaska J.~X.,  Thom C.,  Tumlinson J.,  Tripp T.~M.,  O'Meara
  J.~M.,   Peeples M.~S.,  2013, The Astrophysical Journal Supplement Series,
  204, 17

\bibitem[\protect\citeauthoryear{Wiener, Pfrommer  \& Oh}{Wiener
  et~al.}{2017a}]{Wiener2017}
Wiener J.,  Pfrommer C.,   Oh S.~P.,  2017a, \mn@doi [MNRAS]
  {10.1093/mnras/stx127}, 467, 906

\bibitem[\protect\citeauthoryear{Wiener, Pfrommer  \& Oh}{Wiener
  et~al.}{2017b}]{wiener2017cosmic}
Wiener J.,  Pfrommer C.,   Oh S.~P.,  2017b, Monthly Notices of the Royal
  Astronomical Society, 467, 906

\bibitem[\protect\citeauthoryear{Wiener, Zweibel  \& Oh}{Wiener
  et~al.}{2018}]{wiener2018high}
Wiener J.,  Zweibel E.~G.,   Oh S.~P.,  2018, Monthly Notices of the Royal
  Astronomical Society, 473, 3095

\bibitem[\protect\citeauthoryear{Wiener, Zweibel  \& Ruszkowski}{Wiener
  et~al.}{2019}]{Wiener2019}
Wiener J.,  Zweibel E.~G.,   Ruszkowski M.,  2019, \mn@doi [MNRAS]
  {10.1093/mnras/stz2007}, 489, 205

\bibitem[\protect\citeauthoryear{Xu \& Stone}{Xu \& Stone}{1995}]{Xu1995}
Xu J.,  Stone J.~M.,  1995, The Astrophysical Journal, 454, 172

\bibitem[\protect\citeauthoryear{Yang \& Ruszkowski}{Yang \&
  Ruszkowski}{2017}]{yang2017spatially}
Yang H.-Y.,  Ruszkowski M.,  2017, The Astrophysical Journal, 850, 2

\bibitem[\protect\citeauthoryear{Yang, Ruszkowski, Ricker, Zweibel  \&
  Lee}{Yang et~al.}{2012}]{yang2012fermi}
Yang H.-Y.,  Ruszkowski M.,  Ricker P.,  Zweibel E.,   Lee D.,  2012, The
  Astrophysical Journal, 761, 185

\bibitem[\protect\citeauthoryear{Yang, Ruszkowski  \& Zweibel}{Yang
  et~al.}{2013}]{yang2013fermi}
Yang H.-Y.~K.,  Ruszkowski M.,   Zweibel E.,  2013, Monthly Notices of the
  Royal Astronomical Society, 436, 2734

\bibitem[\protect\citeauthoryear{Yun, Reddy  \& Condon}{Yun
  et~al.}{2001}]{yun2001}
Yun M.~S.,  Reddy N.~A.,   Condon J.~J.,  2001, The Astrophysical Journal, 554,
  803

\bibitem[\protect\citeauthoryear{Zhang et~al.,}{Zhang
  et~al.}{2013}]{zhang2013narrow}
Zhang B.,  et~al., 2013, The Astrophysical Journal, 777, 122

\bibitem[\protect\citeauthoryear{Zweibel}{Zweibel}{2017}]{zweibel2017basis}
Zweibel E.~G.,  2017, Physics of Plasmas, 24, 055402

\bibitem[\protect\citeauthoryear{Zweibel}{Zweibel}{2020}]{zweibel2020role}
Zweibel E.~G.,  2020, The Astrophysical Journal, 890, 67

\makeatother
\end{thebibliography}

%%%%%%%%%%%%%%%%%%%%%%%%%%%%%%%%%%%%%%%%%%%%%%%%%%

%%%%%%%%%%%%%%%%% APPENDICES %%%%%%%%%%%%%%%%%%%%%

\appendix

\section{Convergence}
\label{sec:AppConvergence}
\ryChange{Due to computational expense, we are somewhat limited in physical resolution $\sim$127\,pc in the runs presented above. We expect the accretion rate towards the center of the galaxy to be unresolved and hence we caution interpretation of the central regions of Figure} \ref{fig:Accretion} \ryChange{with a grey region.}

\begin{figure}
  \begin{center}
    \leavevmode
    \includegraphics[width=0.49\textwidth]{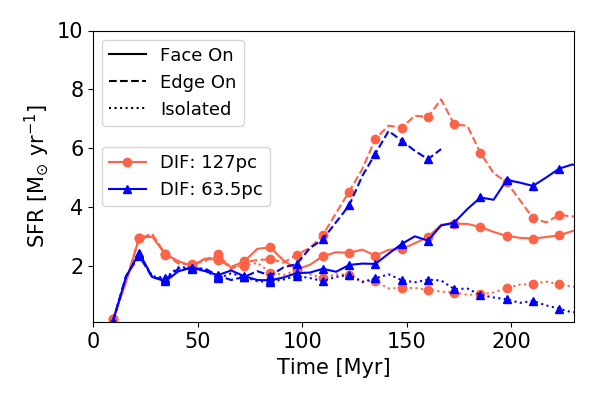}
\caption[]{\ryChange{Time series of SFR for the \FaceOn\ (solid), \EdgeOn\ (dashed) and \Isolated\ (dotted) DIF runs comparing our fiducial 127\,pc resolution (red circles) to a set of high 63.5\,pc resolution (blue triangles) runs. Overall, we find good agreement between the fiducial and high resolution runs.}}
\label{fig:SFH_convergence}
\end{center}
\end{figure}

\ryChange{Although the expense was rather large, we managed to perform simulations of the DIF runs at $\sim$63\,pc resolution to test convergence. Note that the \EdgeOn\ run was the most expensive due to the enhanced star formation rate. Nevertheless, we were able to run the EdgeOn case to the peak in star formation at 175 Myr. We find rather good agreement in the star formation history across resolution, as shown in Figure} \ref{fig:SFH_convergence}. 

\ryChange{Importantly, the results are qualitatively the same: EdgeOn boosts SFRs more than FaceOn stripping. In detail, however, the FaceOn-DIF run at higher resolution attains a stronger burst at > 175 Myr, whereas the Isolated-DIF run at higher resolution exhibits lower SFR. The combined effect is a boost in star formation due to ram pressure stripping beyond the observed range. However, we had similarly disfavored the FaceOn-DIF run based on the accretion rate plot, so our end results are not impacted, even if the details are a bit different. Nevertheless, the SFR is quite similar in comparing the fiducial and high resolution runs for most of the simulated time for most of the physics and wind cases. This is rather surprising as high-resolution ISM slab simulations suggest star formation rates don’t converge until $\sim$8\,pc resolution} \citep[][]{kim2017three}.

\section{True Diode Boundary Conditions}
\label{sec:AppTrueDiodeBCs}
Diode boundary conditions allow material to flow out of the computational domain but are intended to prevent inflow. In FLASH, boundary conditions apply conditions to \textit{cell-centered} values of ghost cells which border the physical computational domain. The default diode boundary condition implementation in FLASH simply replaces the velocity component component normal to the boundary with zero when it is negative. However, since the Riemann problem solves for the fluxes at cell \textit{interfaces} to update hydrodynamic terms, even if the ghost cell-centered velocity is zero, an inflowing velocity at the ultimate interior cell (UIC) will result in an inflow accross the interface. This problem generically leads to non-conservation of mass when gravitational acceleration is present (as well as similar acceleration profiles). 

\begin{figure}
  \begin{center}
    \leavevmode
    \includegraphics[width=0.49\textwidth]{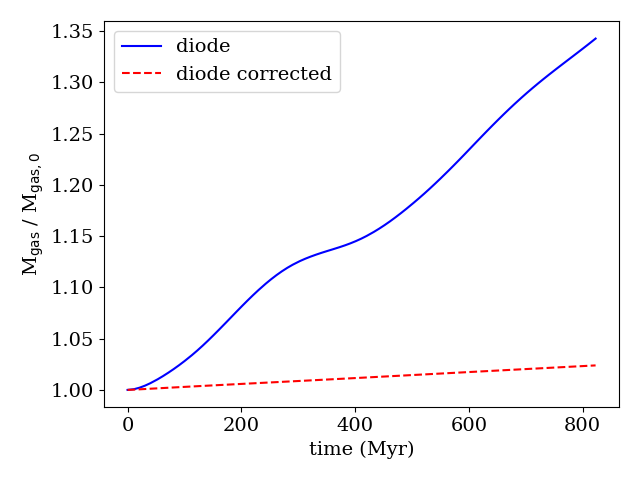}
\caption[]{Conservation of mass with time. The solid blue (dashed red) curves indicate the default (corrected) diode boundary conditions. The corrected diode boundary conditions show marked improvement in mass conservation.}
\label{fig:UIC}
\end{center}
\end{figure}

Therefore, we apply the diode condition to the UIC cells, guaranteeing the flux across the physical domain boundary does not permit inflows. This simple, yet effective modification ensured conservation of mass. In our test, we initialized an $8^3$ box with a static gas of uniform temperature $7 \times 10^7$\,K and density $10^{-27}$g cm$^{-3}$. We turned off magnetic fields, cooling and heating, and star formation and feedback. However, we maintained the static potentials described in Section \ref{subsec:NumericalTechniques}. In Figure \ref{fig:UIC} we show default (corrected) diode boundary conditions with solid blue (red dashed) curves, with the corrected diode boundary conditions indicating marked improvement.

We note that with large box sizes and higher resolution such that the acceleration at the UIC vanishes, non-conservation of mass also vanishes. Related to computational constraints, we were unable to perform our fiducial simulations in larger box sizes (except at lower resolution).

%%%%%%%%%%%%%%%%%%%%%%%%%%%%%%%%%%%%%%%%%%%%%%%%%%

% Don't change these lines
\bsp	% typesetting comment
\label{lastpage}
\end{document}